\documentclass[11pt]{article}
\usepackage{amsmath}
\usepackage{graphicx}
\usepackage{amssymb}

\newcommand{\R}{\mathbb{R}}
\newcommand{\C}{\mathbb{C}}

\normalbaselines
\catcode `\@=11
\@addtoreset{equation}{section}

\def\section{\@startsection {section}{1}{\z@}{-3.5ex plus -1ex minus
     -.2ex}{2.3ex plus .2ex}{\normalsize\bf}}
\def\subsection{\@startsection{subsection}{2}{\z@}{-3.25ex plus -1ex minus
 -.2ex}{1.5ex plus .2ex}{\normalsize\bf}}
\catcode `\@=12
\newcommand{\be}{\begin{equation}}
\newcommand{\ee}{\end{equation}}

\begin{document}

\noindent\textbf{TWISTORS, SPECIAL RELATIVITY, CONFORMAL SYMMETRY AND
MINIMAL COUPLING - A REVIEW.} \vskip35pt \noindent\hspace*{1in} 
\begin{minipage}{13cm}
\bf Andreas Bette, \rm\\ \\
Departamento de Matem{\'a}ticas \\
Divisi{\'o}n de Ciencias Exactas y Naturales\\
Universidad de Sonora, \\  Hermosillo, Mexico. \\ \\
and \\ \\
Royal Institute of Technology, \\
KTH Syd, Campus Telge, \\
S-151 81 S{\"o}dert{\"a}lje, Sweden,\\

\today. \\

e-mail: bette@kth.se or ab@physto.se
\end{minipage}
\vspace*{0.5cm}

\begin{abstract}
\noindent An approach to special relativistic dynamics using the language of
spinors and twistors is presented. Exploiting the natural conformally
invariant symplectic structure of the twistor space, a model is constructed
which describes a relativistic massive, spinning and charged particle,
minimally coupled to an external electro-magnetic field. On the two-twistor
phase space the relativistic Hamiltonian dynamics is generated by a Poincar{%
\'e} scalar function obtained from the classical limit (appropriately
defined by us) of the second order, to an external electro-magnetic field
minimally coupled, Dirac operator. In the so defined relativistic classical
limit there are no Grassman variables. Besides, the arising equation that
describes dynamics of the relativistic spin differs significantly from the
so called Thomas Bergman Michel Telegdi equation.
\end{abstract}
\vspace*{0.5cm}
\section{INTRODUCTION.}
%

\vskip10pt \noindent In relativistic physics one talks often about Lorentz
four-vectors and Lorentz four-tensors. In this introduction we wish to
analyze the mathematical meaning of these notions in a narrative and
formula-free way. Still we would like to keep a certain degree of
strictness. Formulas making our attempts a little more formal will appear in
the next sections.

\vskip10pt \noindent Lorentz four-vectors and Lorentz four-tensors are
vectors and tensors in the usual \emph{vector space} $\R^{4}$ equipped with
the pseudo-Euclidean (Minkowski) metric $\eta$ of signature $(+---)$.
Besides of being vectors and tensors in $\R^{4}$, they are also invariant
geometrical objects with respect to the action of the homogeneous Lorentz
group preserving $\eta$. It is very well-known that, because of the
pseudo-Euclidicity of $\eta$, the vectors in the vector space $\R^{4}$ are
divided into three types: time-like (their Minkowki norms are positive),
space-like (their Minkowski norms are negative) and null-like (their
Minkowski norms are zero). The null-like vectors are very special and, as
mentioned above, are invariant geometrical objects with respect to the
homogeneuos Lorentz group. However they are also invariant with respect to a
change of scale, i.e. \@ prolongations or shortenings of a null-like vector
does not change its vanishing Minkowski norm.

\vskip10pt \noindent The connected component of the identity of the homogeneous
Lorentz group, regarded as a six dimensional smooth manifold, is not simply
connected but covered by a group (simply connected six dimensional smooth
manifold) that can be represented by the matrix group\footnote{$2 \times2$
matrices with complex entries and determinants equal to one.} $\mathrm{SL}(2,\C)$. The
connected component of the identity of the homogeneous Lorentz group becomes just a
two to one homomorphic image of $\mathrm{SL}(2,\C)$. Two matrices in $\mathrm{SL}(2,\C)$, one
with plus and one with minus sign, represent the same element of the
identity connected component of the homogeneous Lorentz group.

\vskip10pt \noindent$\mathrm{SL}(2,\C)$ matrices act naturally on complex vectors in $%
\C^{2}$. The group $\mathrm{SL}(2,\C)$ preserves an antisymmetric form (a ``metric'') $%
\epsilon$ in $\C^{2}$. The complex vectors in $\C^{2}$ regarded as geometrical
objects with respect to $\epsilon$ are called (Weyl) spinors and $\C^{2}$
equipped with $\epsilon$ is called the (Weyl) spinor space $S$. An entire $%
\mathrm{SL}(2,\C)$ invariant tensor algebra over $S$ (and its complex conjugate
counterpart) arises in this way.

\vskip10pt \noindent It turns out then that null-vectors in the Minkowski
vector space may be regarded as simple hermitian spinor tensors of second
rank in this tensor algebra over $S$ and its complex conjugate, i.e.\@ each
spinor together with its complex conjugate counterpart defines a
null-vector. More exactly each null-vector in Minkowski vector space is
defined by a unique set of spinors in $S$ given by a spinor modulo its
multiplication by a complex number whose absolute value is equal to one
i.e.\@ modulo multiplication by a phase factor. Although spinors are
simplest geometrical objects (complex vectors) in $S$ their interpretation
in terms of geometrical objects in the Minkowski vector space is not so
simple. For example, the phase factor (angle) of a spinor has an exact but
quite complicated geometrical meaning\footnote{%
it describes an angle variable associated with the so called flag of a
spinor, see developments in the next sextion and e.g.\@\cite{prrw}.} in the
Minkowski vector space. Spinors modulo multiplication by a non-zero
complex number\footnote{%
in other words complex lines through the origin in $\C^{2}$ defining all
points on $\C \mathrm{P}(1)$ (complex projective space of dimension one), i.e.\@ on the
Riemann sphere.} represent null-direction in the Minkowski vector space,
i.e.\@ represent null-vectors modulo their multiplication by a non-zero
real number. This and much much more is carefully described in, for example,
Penrose's and Rindler's book \cite{prrw}.

\vskip10pt \noindent In this review, Lorentz four-vectors and four-tensors
(and thereby Minkowski vector space itself) will be treated as a subset of
the ($\mathrm{SL}(2,\C)$ invariant) spinor tensor algebra over $S$ and its complex
conjugated counterpart. Minkowski vector space is thus regarded as less
elementary than $\C^{2}$, the complex vector space equipped with the
``metric'' $\epsilon$, see \cite{Stewart} chapter 2 and Appendix A.

\vskip10pt \noindent One of the remarkable insights that follows from such
point of view and nowadays used extensively in calculations within general
relativity (see for example, \cite{pr1,prrw,Stewart}), is that any orthonormal basis
(three space-like directions and one time-like direction) with respect to
the pseudo-Euclidean metric $(+---)$ in the Minkowski vector space may be
regarded as constructed out of just any pair of non-parallell
(non-proportional) spinors (i.e. a pair of spinors with non-vanishing
``scalar product'' with respect to the $\epsilon$ ``metric'') that are
normalised to one (two such spinors are said to form a spin frame. Note that 
$\epsilon$-norm of any spinor is always equal to zero).

\vskip10pt \noindent More concrete developments will follow in the sequel.
But first let us make some remarks on the \emph{difference} between the 
\emph{Minkowski vector space} as referred to above and the \emph{affine
Minkowski space} representing the space of physical events. Let us then
discuss briefly how this might lead us, in a natural way, to the
introduction of the twistor space.

%

\vskip10pt \noindent The Lorentz tensor algebra over the \emph{Minkowski vector
space} is not sufficient for the description of (special) relativistic
physical phenomena. An upgrading of the Minkowski vector space to the \emph{%
affine Minkowski space} is needed. The points in the affine Minkowski space
represent a postulated geometrical continuum (?) of physical events. Events
are not Lorentz four-vectors although they may be identified and represented
by \emph{position} four-vectors. It is essential to be aware of the
difference between these two geometrical objects. In practical applications
an origin is always chosen to start with, so this distinction is not always
so transparent. On the other hand, the four-intervals between two arbitrary
events in space-time are, of course, Lorentz four-vectors.

\vskip10pt \noindent The invariance group of the affine Minkowski space
(compared with the Lorentz group invariance of just the Minkowski
four-vector space) is now extended by the arbitrariness of the choice of the
space-time origin. The composition of the homogeneous Lorentz group with the
group of translations of the origin in space-time\footnote{%
note that these two subgroups, i.e.\@ the translations and the Lorentz
transformations do not commute in general.} is called the Poincar{\'e} group
or sometimes the inhomogeneous Lorentz group\footnote{%
while the Lorentz group manifold is six dimensional (six parameters), the
Poincar{\'e} group manifold is ten dimensional.}. At each point (event) in
the affine Minkowski space, the space of the Lorentz invariant
four-intervals pointing towards all other points of the affine Minkowski
space form a Minkowski vector space, i.e.\@ they form a Minkowski tangent
space at that point. As mentioned above, to label an event in space-time a
new notion is needed, namely, the notion of a \emph{position four-vector}.
Such position four-vectors behave like usual four-vectors under the action
of the Lorentz group but in contrast to genuine four-vectors, they are also
affected by changes of the space-time origin, i.e.\@ by space-time
translations.

\vskip10pt \noindent By analogy with the special role played by
null-vectors in the Minkowski vector space (effectively only one spinor
modulo its phase defines such a null-vector) straight null-lines in the
Minkowski space play also an exceptional role. These null-lines may be
identified with possible trajectories of free massless and spinless
particles carrying linear and angular (with respect to some arbitrarily
chosen space-time origin) four-momenta. The pseudo-Euclidean norms of
four-intervals between any two arbitrary space-time points along such a
null-line equal zero. This implies that multiplying genuine null-vectors at
each space-time point by a non-zero real number (different at each
space-time point) does not affect the null-lines (passing through this point
and) having these null-vectors as directions at this point. In order to keep
the straight null-lines in Minkowski space unaffected it is therefore
sufficient to require that the Minkowski metric is preserved only modulo its
multiplication by an arbitrary non zero positive real valued function on
the Minkowski space\footnote{%
it may be shown that in the Minkowski space this is equivalent to the
requirement that the Lorentz scalar product of any two non-null Minkowski
intervals divided by the product of their norms should be unaffected hence
the name conformal invariance; ``the Minkowski angles'' are to be preserved.}%
. Thus the set of all straight null-lines in the Minkowski space is
invariant with respect to an even larger symmetry group than than the
symmetry (Poincar{\'e}) group of the Minkowski space itself. A (non-linear
and local) representation in the Minkowski space itself of this enlarged
fifteen dimensional conformal symmetry group has been known for quite a long
time and is denoted by $C(1,3)$. The first ten dimensions of its Poincar{\'e}
subgroup (manifold) have a clear physical meaning. They represent Lorentz
transformations (rotations and the so called boosts) and the
four-translations. The fundamental laws of nature, stating that the angular
momentum and the velocity of the centre of the energy of an isolated
physical system are always conserved, may be derived from the Lorentz
symmetry while the system's conservation of its energy and its linear
momentum, may be derived from the translational symmetry. The remaining
symmetry concealed in the five additional parameters of the conformal group
has (in terms of conservation laws) no clear physical meaning at the moment.
There is a lot of subtle points concerning this enlarged (conformal)
invariance group. Many of them will not be discussed here, for example, we
will not discuss the fact that the action of this enlarged invariance
conformal group is non-linear and only local while represented in the
Minkowski space and what this possibly implies, etc. The reader may consult
original papers on this topic \cite{jmp,prmc,prrw} (and references therein)
for a detailed discussion of the issue. In this review we will focus
only on certain aspects that will lead us into models describing dynamics of
relativistic spinning massive particles in terms of the so called twistors.

\vskip10pt \noindent What are twistors? They are related to the conformal
group $C(1,3)$ mentioned above in the following fashion. A linear and also
four to one covering group (manifold) of the group $C(1,3)$ may be
represented by the matrix group\footnote{%
special (determinats equal to one) unitary matrices with complex valued
entries preserving the pseudo-hermitian form $g$ of signature $++--$.} $%
SU(2,2)$. The set of complex valued vectors in $\C^{4}$ equipped with a
pseudo-hermitian form $g$ preserved by the action of the $SU(2,2)$ group%
\footnote{%
in order to see explicitly and keep track of how the (spinorial versions of)
the Poincar{\'e} and Lorentz groups are ``inbedded'' inside the $SU(2,2)$ as
subgroups it is necessary to use special spinor representations of $g$ and $%
SU(2,2)$ . See developments in the sequel.} is called the twistor space $T$
of non-projective twistors $\{Z, W, ..\}$. From this it should follow that (the
covering (manifold) of the connected component of the identity of) the Poincar{\'e}
group is a subgroup of the $SU(2,2)$ group.

\vskip10pt \noindent It is then possible to identify certain $SU(2,2)$
and/or Poincar{\'e} invariant/covariant functions on $T$ with geometrical
and dynamical/kinematical variables of a (classical limit of a) massless
spinning object in the Minkowski space. Using at least two copies of $T$,
certain $SU(2,2)$ and/or Poincar{\'e} invariant/covariant geometrical and
dynamical/kinematical variables (including the position four-vectors) of a
(classical limit of a) charged massive spinning object in the Minkowski
space may be also identified.

\vskip10pt \noindent The imaginary part of the pseudo-hermitian form $g$ in $%
\C^{4}$ constitutes an $SU(2,2)$ invariant symplectic structure on $T$ which
allows the non-projective twistor space to be treated as the simplest
possible (extended) phase space of a (classical limit of a) massless
spinning object equipped with globally defined and canonically conjugated
conformally and thereby also Poincar{\'e} invariant/covariant variables.

\vskip10pt \noindent The subset of twistors (modulo multiplication by
non-zero complex numbers with their absolute value equal to one, i.e.\@
modulo multiplication by phase factors), in the twistor space $T \simeq
\C^{4}$, with their pseudo hermitian norms equal to zero (null-twistors) may,
when interpreted in the Minkowski space, be identified with the set of all
possible straight trajectories of massless spinless particles with given
linear null- and (null-) orbital angular four-momenta that with respect to
the above mentioned symplectic structure fulfill the Poisson bracket
commutation relations of the Poincar{\'e} algebra. If the pseudo-hermitian
norm is not equal to zero, then the corresponding twistors may still be
identified with massless spinning objects in the Minkowski space with given
linear null- and (null-) angular four-momenta that again fulfill the
commutation relations of the Poisson bracket Poincar{\'e} algebra with
respect to the symplectic structure defined by the imaginary part of the
pseudo-hermitian form. Remarkably, such (non-quantum, i.e.\@ classical limits
of) massless spinning objects do not have well-defined trajectories when one
tries to interpret them in the (real) Minkowski space\footnote{%
instead of trajectories they may be interpreted as twisting Robinson
congruences of null-lines filling up the entire Minkowski space. When the
twist (i.e.\@ the norm of the corresponding twistor) vanishes such a
congruence will meet, and thereby, in fact, define a unique null-line
previously already identified as the null-line represented by this
null-twistor \cite{pr3,jmp,prmc,prrw}.}. However, taking into account the
phase space structure of $T$, pairs of non-coinciding twistors (or quite
generally any number (greater than two) of non-coinciding twistors) may also
be used to define dynamical variables representing four-positions \cite
{zab,zab1,hl,prmc} of charged massive and, in general, spinning object in the
Minkowski space. This remarkable fact\footnote{%
remarkable because non-local massless objects in Minkowski space define
explicitly an event (extremely local object) in space-time. Events (local
objects) become secondary, while non-local massless objects (in general
represented by Robinson congruences) are primary in such an approach.} will
be utilised extensively by us in this review.

\vskip10pt \noindent Summarizing, the description of (massless non-local)
geometrical objects, ``living'' in the Poincar{\'e} invariant Minkowski
space, in terms of abstract conformally invariant (local) geometrical
objects ``living'' in the twistor space, uses the fact that the conformal
group representation $C(1,3)$ of the Minkowski space is a 4-1 homomorphic
image of the complex matrix group $SU(2,2)$ acting on the twistor space $T
\simeq \C^{4}$. Any element in $C(1,3)$ may be represented by a matrix $A$ in 
$SU(2,2)$ or by $-A$ or by $iA$ or finally by $-iA$. $C(1,3)$ is thus a
homomorphic image of $SU(2,2)$. Twistors with norms equal to zero are
identified with null-lines in the Minkowski space. Non-null twistors are
geometrically identified with Robinson congruences. Any geometrical object
``living'' in the Minkowski space may, in a relatively easy way, be described in
terms of geometrical objects (twistor-tensors) in the twistor space but not
necessarily, and not so easily, in the opposite direction.

\vskip10pt \noindent To be somewhat more exact in the definition of twistors
let us make certain additional remarks, in risk of being repetitive of what already has
been said above. The complex vectors in $\C^{4}$ equipped with a ``metric''
(pseudo-hermitian form $g$) preserving $SU(2,2)$ are, in fact, called \emph{%
non-projective} twistors. The pseudo-hermiticity implies that the norm of a
non-projective twistor in $T$ may assume positive, null or negative real
values. Non-projective twistors having their norms equal to zero are called
non-projective, null-twistors as already mentioned. The non-projective
twistors with positive norms are called non-projective, positive helicity
twistors. The non-projective twistors in $T$ with negative norms are called
non-projective, negative helicity twistors. Non-projective twistors modulo
multiplication by non-zero complex numbers (changing the value of its norm
but not the sign) are called \emph{projective} twistors and form a space of
complex lines through the origin in the complex vector space $\C^{4}$. The
set of complex lines in $\C^{4}$ forms a six dimensonal manifold (with
topology $S^{7}/S^{1}$) and is denoted by $\C P(3)$. Note that the pseudo-Euclidean
norm in $T$ splits $\C P(3)$ is into three parts depending on the
sign $(+,0,-)$ of the corresponding non-projective twistor representatives. Thus, $%
T$ is the space of homogeneous coordinates of $\C P(3)$ or, if one so
wishes to identify it, the complex line bundle over $\C P(3)$.

\vskip10pt \noindent It has been shown by us \cite{zab,zab1,zak,zak2,zak1}
that twistors and the symplectic structure defined by the imaginary part of
the pseudo-hermitian form preserved by $SU(2,2)$ define an extended phase
space of massive, electrically charged and in general spinning objects.
Among the relativistic dynamical/kinematical variables describing these
objects, the Minkowskian space-time positions are singled out by the
formalism. In the simplest case, the four-positions of the objects are \emph{%
identified} as certain Poincar{\'e} covariant functions of two
non-coinciding twistors. Physical events become in this way (at least
mathematically) secondary objects, \emph{defined} by twistor variables. This
way of looking at space-time events is analogous (but somewhat more subtle)
to the way one regards any time-like or space-like Lorentz four-vector as
constructed from a pair of non-proportional spinors in the spinor space (see
e.g.\@\cite{Stewart} chapter 2 and Appendix A).

\vskip10pt \noindent The relativistic dynamics of a (classical limit of a)
massive and spinning particle-object as mentioned above, may be viewed as
a canonical flow generated by an appropriately chosen real valued function
(e.g.\@ by identifying the classical limit of the second order minimally
coupled Dirac operator as such a function) on a direct product of two copies%
\footnote{%
less the diagonal points in that product.} of $T$ where the relativistically
invariant canonical symplectic structure consists of a direct sum of two
copies of the imaginary parts of the pseudo-hermitian form preserved by the
action of $SU(2,2)$ \cite{ab5,ab1}. We now proceed to make all the
statements above more concrete, mathematically. We will be very
sloppy with proofs because we wish only to present the known results and use
them promptly for our purposes, namely, in order to derive from the formalism
the relativistic dynamical equations describing a massive, charged and
spinning ``particle'' moving in an external electro-magnetic field.


\section{SPINORS, LORENTZ FOUR-VECTORS AND LORENTZ\\
FOUR-TENSORS.}

In this section we start with the (abstract) vector space $\C^{2}$ equipped
with a $\mathrm{SL}(2,\C)$ invariant antisymmeric ``metric'' $\epsilon$. Such a
two-dimensional complex vector space is called the Weyl spinor\footnote{%
we will in most cases just write ``spinor'' omitting the name of Weyl.}
space $S$ and will be used to construct the physical (less abstract) notion
of the Minkowski vector space $M_{v}$, i.e.\@, a real vector space $\R^{4}$
equipped with the Minkowski metric $\eta$ which is invariant with respect to
the $SO(1,3)$ group representing the (identity connected) homogeneous
Lorentz group. As a consequence, the Lorentz tensor algebra will
appear as a subset of the complex valued spinor-tensor algebra (see e.g.\@
\cite{pr1,prrw}).

\vskip10pt \noindent We use greek lower case letters to denote spinors and
spinor-tensors in $S$. The greek lower case letter $\epsilon$ will be
reserved to denote the $\mathrm{SL}(2,\C)$ invariant antisymmetric ``metric'' tensor
in $\C^{2}$. Note that for each spinor in $S$ there exists its complex
conjugate counterpart. Besides, for each spinor there exists also its
covariant counterpart (with respect to the $\mathrm{SL}(2,\C)$ (${\overline{\mathrm{SL}(2,\C)}}$%
) invariant antisymmetric ``metric'' $\epsilon$ (${\bar\epsilon}$)). Thus,
each contravariant spinor has three ``brothers'', its covariant version, its
complex conjugate, and the covariant version of its complex conjugate. We
will distinguish among them according to the following well-known
convention: the contravariant spinors will be distinguished by latin upper
case superscript letters taking values 0 and 1 with respect to some spin
frame (i.e.\@ a normalised spinor basis). If no frame is chosen, the index
just tells us what kind of entity we are dealing with (Penrose's abstract
index notation \cite{pr1}). The covariant spinors will be distinguished by
latin upper case subscript letters taking values 0 and 1 with respect to
such a spin frame. Complex conjugation will be denoted by a bar over the
symbol with simultaneous priming of the subscript and superscript letters.
According to this convention we will denote the spinor space and the
corresponding complex conjugate spinor space by

\begin{equation}
{S}=(\C^{2}, \ \epsilon_{AB}), \ \ \ {\bar S}=({\bar \C}^{2}, \ {\bar\epsilon }%
_{A^{\prime}B^{\prime}}). \label{spinorspace}
\end{equation}

\vskip10pt \noindent In addition, we will have the covariant versions of
these spaces that will be denoted by

\begin{equation}
{S^{*}}=(\C^{2}, \ \epsilon^{AB}), \ \ \ {{\bar S}^{*}}=({\bar \C}^{2}, \ {%
\bar\epsilon}^{A^{\prime}B^{\prime}}), \label{spinorspacestar}
\end{equation}

\vskip10pt \noindent where $\epsilon^{AB}$ is the inverse of $\epsilon_{BA}$
while ${\bar\epsilon}^{A^{\prime}B^{\prime}}$ is the inverse of ${\bar
\epsilon}_{B^{\prime}A^{\prime}}$\footnote{%
so that we always have $\epsilon^{CB} \epsilon_{BD}=\delta^{C}_{D}$, i.e.\@
mappings from $S$ to $S^{*}$ (or from ${\bar S}$ to ${\bar S}^{*}$)
correspond to a lowering of the ``contravariant'' spinor index towards the
``covariant'' index nearest to the kernel letter $\epsilon$ (or ${\bar{%
\epsilon}}$) while the inverse mapping is a raising of a ``covariant''
spinor index toward the second contravariant index in the kernel letter $%
\epsilon$ (or ${\bar{\epsilon}}$), a good mnemonic rule.}. The invariance of
the $\epsilon$-``metric'' may now be expressed as follows

\begin{equation}
\epsilon_{AB} = L^{C}_{A} \ L^{D}_{B} \ \epsilon_{CD}, \ \ \ \ \ {\bar
\epsilon}_{A^{\prime}B^{\prime}}= {\bar L}^{C^{\prime}}_{A^{\prime}} \ {\bar
L}^{D^{\prime}}_{B^{\prime}} \ {\bar\epsilon}_{C^{\prime}D^{\prime}},  \label{inv}
\end{equation}

\vskip10pt \noindent where 
\begin{equation}
L^{C}_{A} \in \mathrm{SL}(2,\C) \ \ \ {\bar L}^{C^{\prime}}_{A^{\prime}} \in \overline{%
SL(2,{\C})}. \label{LinSL2}
\end{equation}

\vskip10pt \noindent With respect to a spin frame in $\C^{2}$ we may use the
following numerical representations of the antisymmetric ``metric''
preserved by the action of $\mathrm{SL}(2,\C)$ and of $\overline{SL(2,{\C})}$ on $%
\C^{2}$ and on its complex conjugate ${\bar \C}^{2}$:

\begin{equation}
\epsilon_{AB} = \left( 
\begin{array}{cc}
0 & 1 \\ 
-1 & 0
\end{array}
\right) , \ \ \ \ {\bar\epsilon}_{A^{\prime}B^{\prime}} = \left( 
\begin{array}{cc}
0 & 1 \\ 
-1 & 0
\end{array}
\right) , \label{epsilon}
\end{equation}

\begin{equation}
\epsilon^{AB} = \left( 
\begin{array}{cc}
0 & 1 \\ 
-1 & 0
\end{array}
\right) , \ \ \ \ {\bar\epsilon}^{A^{\prime}B^{\prime}}= \left( 
\begin{array}{cc}
0 & 1 \\ 
-1 & 0
\end{array}
\right) . \label{epsilonbar}
\end{equation}

\vskip10pt \noindent Thus, the antisymmetric ``metrics'' $\epsilon$ and $%
\bar\epsilon$ preserved by $\mathrm{SL}(2,\C)$ and ${\overline{\mathrm{SL}(2,\C)}}$ are also
symplectic structures on $\C^{2}$ and on ${\bar \C}^{2}$ respectively.
Consequently the ``norm'' of any spinor is, by definition, always equal to
zero ($\epsilon_{AB} \pi^{A} \pi^{B}=0$ for any $\pi$ in $S$).

\vskip10pt \noindent Any element in $SL(2, \C)$ and in its complex conjugate $%
{\overline{SL(2,{\C})}}$ is, with respect to a spin frame, represented by
the following set of two by two matrices with complex entries, matrices
having their determinants equal to one\footnote{%
the four complex valued entries in the matrix will then represent just six
real variables namely the six continuous parameters of the Lorentz group
which will be seen more clearly later on in this review.}: \vskip10pt
\noindent

\begin{equation}
L^{M}_{N} = \left( 
\begin{array}{cc}
a & b \\ 
c & d
\end{array}
\right) \in SL(2, \C), \label{L}
\end{equation}

\begin{equation}
{\bar L}_{N^{\prime}}^{M^{\prime}}= \left( 
\begin{array}{cc}
{\bar a} & {\bar b} \\ 
{\bar c} & {\bar d}
\end{array}
\right) \in{\overline{SL(2, {\C})}}, \label{Lconjugate}
\end{equation}

\begin{equation}
l^{N}_{M} := (L^{-1})^{N}_{M} = \left( 
\begin{array}{cc}
d & -b \\ 
-c & a
\end{array}
\right) \in SL(2, \C), \label{Linverse}
\end{equation}

\begin{equation}
{\bar l}_{M^{\prime}}^{N^{\prime}}:=({\bar L}^{-1})^{N^{\prime}}_{M{^{\prime}%
}} = \left( 
\begin{array}{cc}
{\bar d} & {- \bar b} \\ 
{- \bar c} & {\bar a}
\end{array}
\right) \in{\overline{SL(2, {\C})}}. \label{Linverseconjugate}
\end{equation}

\vskip10pt \noindent The invariance of the $\epsilon$-``metric'' in (\ref{inv})\
may be now, with respect to a spin frame, expressed in matrix form as:

\begin{equation}
L^{T} \epsilon L = \epsilon, \ \ \ {\bar L}^{T} {\bar\epsilon} {\bar L} = {%
\bar\epsilon}, \label{Lmatrix}
\end{equation}

\vskip10pt \noindent where the superscript $T$ denotes transposition of a
matrix.

\vskip10pt \noindent A simple exercise is to verify identities in (\ref{Lmatrix}) by using 
the matrix representaions in (%
\ref{epsilon})--(\ref{Lconjugate}).

\vskip10pt \noindent According to the above described convention, whether
the spin frame is chosen or not, any contravariant spinor, its covariant
version and their complex conjugate counterparts may be explicitly written
as

\begin{equation}
\pi^{A} \in S, \ \ \ \pi_{B}:=\pi^{A} \epsilon_{ AB} \in S^{*}, \ \ \ {\bar
\pi}^{A^{\prime}} \in{\bar S}, \ \ \ {\bar\pi}_{B^{\prime}}:= {\bar\pi }%
^{A^{\prime}}{\bar\epsilon}_{A^{\prime}B^{\prime}} \in{\bar S}^{*}, \label{spinorbrothers}
\end{equation}

\vskip10pt \noindent and conversely:

\begin{equation}
\pi^{A} = \pi_{B} \epsilon^{AB} \in S, \ \ \ {\bar\pi}^{A^{\prime}}:= {\bar
\pi}_{B^{\prime}}{\bar\epsilon}^{A^{\prime}B^{\prime}} \in{\bar S}. \label{spinorbrothers1}
\end{equation}

\vskip10pt \noindent Note the order of contractions, i.e.\@ to obtain a
covariant spinor contraction with the first $\epsilon$ subscript letter is
performed, while a contravariant spinor is created by contraction with the
second superscript letter of the $\epsilon$-``metric''.

\vskip10pt \noindent Obviously, action of an $\mathrm{SL}(2,\C)$ transformation on a
spinor reads

\begin{equation}
{\widetilde\pi}^{M}= L^{M}_{N} \pi^{N} , \ \ \ {\widetilde{\bar\pi}}%
^{M^{\prime}}={\bar L}_{N^{\prime}}^{M^{\prime}} {\bar\pi}^{N^{\prime}} , \
\ \ {\widetilde\pi}_{M}= l^{N}_{M} \pi_{N} , \ \ \ {\widetilde{\bar\pi}}%
_{M^{\prime}}={\bar l}_{M^{\prime}}^{N^{\prime}} {\bar\pi}_{N^{\prime}}. 
                                            \label{spinortransformations}
\end{equation}

\vskip10pt \noindent Consider now all hermitian ($v^{AA^{\prime}} ={\bar v}%
^{A^{\prime}A}$) mixed (i.e., once unprimed and once primed) spinor-tensors
of second rank given by

\begin{equation}
v^{AA^{\prime}}:=a \pi^{A} {\bar\pi}^{A^{\prime}} + u \pi^{A} {\bar\eta }%
^{A^{\prime}} + {\bar u} \eta^{A} {\bar\pi}^{A^{\prime}} + b \eta^{A} {%
\bar\eta}^{A^{\prime}},  \label{hermitiansecond}
\end{equation}

\vskip10pt \noindent where $\pi$ and $\eta$ are two non-proportional spinors
(i.e.\@$\epsilon_{AB} \pi^{A} \eta^{B} \not = 0$), $a$, $b$ arbitrary real
numbers and $u$ an arbitrary complex number. The set of such hermitian mixed
spinor-tensors will be denoted by $(S \otimes{\bar S})_{h} $. The set of all
mixed spinor-tensors of rank two\footnote{%
with $a, \ b$ in (\ref{hermitiansecond}) being now arbitrary complex number
and the complex numbers $u$, $\bar u$ not being necessarily complex
conjugate to each other.} will be denoted by $S \otimes{\bar S}$.

\vskip10pt \noindent From (\ref{spinortransformations}) it now follows that
any mixed spinor-tensor of second rank and therefore, any hermitian one,
under the action of an $\mathrm{SL}(2,\C)$ transformation, is transformed as follows

\begin{equation}
{\widetilde v}^{AA^{\prime}}=L^{A}_{B} {\bar L}_{B^{\prime}}^{A^{\prime}}
v^{BB^{\prime}}. \label{Minkowskivector}
\end{equation}

\vskip10pt \noindent The subgroup ($(SL(2,\C) \otimes{\overline{\mathrm{SL}(2,\C)}}%
)_{h}$ of hermitian transformations in $SL(2,\C) \otimes{\overline{\mathrm{SL}(2,\C)}}$
of the type $L^{A}_{B} {\bar L}_{A^{\prime}}^{B^{\prime}}$ as in (\ref
{Minkowskivector}) acting in the space of mixed spinor-tensors of second
rank such as in ($\ref{hermitiansecond}$) is, as well-known, {\em isomorphic} to
the group\footnote{%
preserving the metric $\eta$ of signature $(+---)$ in $\R^{4}$.} $SO(1,3)$.
By definition, the group of hermitian transformations in $SL(2,\C) \otimes{%
\overline{\mathrm{SL}(2,\C)}}$ preserves the hermitian tensor product of the $\epsilon$%
-metrics,

\begin{equation}
\eta_{AA^{\prime}BB^{\prime}} = L^{M}_{A} {\bar L}_{A^{\prime}}^{M^{\prime}}
L^{K}_{A} {\bar L}_{B^{\prime}}^{K^{\prime}} \eta_{MM^{\prime}KK^{\prime}}, \label{invariance}
\end{equation}

\vskip10pt \noindent where

\begin{equation}
\eta_{AA^{\prime}BB^{\prime}}:= \epsilon_{AB} {\bar\epsilon}_{A^{\prime
}B^{\prime}}.         \label{eta}
\end{equation}

\vskip10pt \noindent Because of the above mentioned isomorphism we may
identify the group of hermitian transformations $(SL(2,\C) \otimes {%
\overline{\mathrm{SL}(2,\C)}})_{h}$ of the type displayed in (\ref{Minkowskivector})
with the group $SO(1,3)$ representing the identity component of the
homogenous Lorentz group. This, in turn, implies that $M_{v}$ may be
identified with $(S \otimes{\bar S})_{h} $ and that $\epsilon{\bar\epsilon}$
may be identified with the Minkowski metric $\eta$ as displayed in (\ref{eta}%
). For each (abstract) Lorentz four-vector $v^{a}$ we thus have the
identification

\begin{equation}
v^{a}=v^{AA^{\prime}}={\bar v}^{A^{\prime}A}. \label{fourvector}
\end{equation}

\vskip10pt \noindent For the Lorentz transformations in $M_{v}$, the above
mentioned isomorphism implies the following identifications

\begin{equation}
L^{m}_{n}=L^{MM^{\prime}}_{NN^{\prime}}=L^{M}_{N} {\bar L}%
_{N^{\prime}}^{M^{\prime}}, \ \ \ ({L^{-1}})^{n}_{m}:=l^{N}_{M} {\bar l}%
_{M^{\prime}}^{N^{\prime}} \ \ \ {\text{where}} \ \ \ L^{m}_{n} \in SO(1, 3). \label{lorentz}
\end{equation}

\vskip10pt \noindent The Minkowski metric $\eta$ now appears as a fourth
order hermitian tensor product of the $\epsilon$ ``metric'' and $\bar{%
\epsilon}$-``metric'' (see (\ref{eta})),

\begin{equation}
\eta_{ab}=\eta_{AA^{\prime}BB^{\prime}}. \label{Minkowskimetric}
\end{equation}

\vskip10pt \noindent Referring to the above identification, one could say
that the symplectic structures $\epsilon$ and $\bar\epsilon$ on the two
spinor spaces $S$ and $\bar S$ represent ``square roots'' of the Minkowski
metric $\eta$ while the two spinor spaces themselves represent ``square
roots'' of the $M_{v}$ (Minkowski \emph{vector} space) itself.

\vskip10pt \noindent It should be understood that exact relations rendering
coordinates of Minkowski four-tensors in terms of coordinates of spinor
tensors\footnote{%
coordinates require, of course, a choice of a spin frame (normalised spinor
basis) in the spinor space $S$.} depend on the choice of the explicit
isomorphism $\sigma^{a}_{AA^{\prime}} \ \sigma^{b}_{BB^{\prime}}$ between $%
(SL(2,\C) \otimes{\overline{\mathrm{SL}(2,\C)}})_{h}$ and $SO(1,3)$, where $%
\sigma^{a}_{AA^{\prime}}$ defines an explicit bijective mapping between mixed
contravariant spinors of second rank and contravariant Lorentz
vectors\footnote{%
possibly complexified if the mixed spinor tensor of second rank is not
hermitian.}. The isomorphism represented by $\sigma^{a}_{AA^{\prime}}$ is
sometimes called the Infeld-Van der Waerden connecting quantity. With respect to
a spin frame in the spin vector space and with respect to an ortho-normal
frame in the Minkowski vector space, the four two-dimensional matrices $%
\sigma^{a}_{AA^{\prime}}$ are simply given by the two-dimensional (chirally
reduced) Dirac matrices.

\vskip10pt \noindent A convenient and frequently used explicit choice of the $%
\sigma^{a}_{AA^{\prime}}$ matrices is provided by three, two by two,
dimensional Pauli matrices and one two by two dimensional identity matrix,
all four matrices multiplied by $\frac{1}{\sqrt{2}}$. This extra numerical
factor is inserted in order to harmonize normalisation of any pair of
non-parallell spinors with respect to the ``metric'' $\epsilon$ (and/or $%
\bar\epsilon$) with ortho-normalisation of the corresponding Lorentz
four-vectors (see (\ref{hermitiansecond})) with respect to the arising
pseudo-Euclidean metric $\eta$ in (\ref{Minkowskimetric}) in $M_{v}$.
However, in this review we shall not need explicit coordinate expressions
very often. All necessary details we are omitting here may be found in
Penrose's and Rindler's book \cite{prrw} (see also \cite{Stewart} chapter 2).

%

%

\vskip10pt \noindent While defining Lorentz four-tensors of various ranks in
terms of hermitian spinor-tensors, it is extremly useful to note that only
the symmetric part of non-mixed (i.e.\@ either primed or non-primed)
spinor-tensors is important, the antisymmetric part reduces itself to
an $\mathrm{SL}(2,\C)$ (or ${\overline{\mathrm{SL}(2,\C)}}$) spinor-tensor contraction times the 
${\epsilon}$ (or ${\bar{\epsilon}}$) ``metric''. This is a consequence of
the almost obvious (Fierz) identity:

\begin{equation}
\pi_{A} \eta_{B} - \pi_{B} \eta_{A} = \epsilon_{CD} \pi^{C} \eta^{D} =
\pi_{D} \eta^{D} \epsilon_{AB}.   \label{Fierz}
\end{equation}

\vskip10pt \noindent Another important property of the spinor-tensor algebra
is that any symmetric spinor tensor is always simple, i.e., of the form

\begin{equation}
\nu^{(ABCD \ldots F)}=\alpha^{(A} \beta^{B} \gamma^{C} \zeta^{D} \ldots
\iota^{F)},   \label{symmetricspinor}
\end{equation}

\vskip10pt \noindent where round brackets, as usual, denote symmetrisation
and the spinors defining $\nu$ in ({\ref{symmetricspinor}) are only unique
up to multiplication by a constant nonzero complex number. Therefore, each such
spinor defines a null-direction (not an entire null four-vector)
in the Minkowski vector-space. Some of these directions may coincide. In this case,
one says that $\nu$ is algebraically special. }

\vskip10pt \noindent For example, the identity in (\ref{Fierz}) and the
property described in (\ref{symmetricspinor}) imply that the antisymmetric
Lorentz four-tensor of second rank, expressed as an hermitian spinor-tensor
of fourth rank, may be written as follows

\begin{equation}
M^{ab} =-M^{ba}= M^{AA^{\prime}BB^{\prime}} = -M^{BB^{\prime}AA^{\prime}}=
\mu^{(AB)} \ \epsilon^{A^{\prime} B^{\prime} } + {\bar\mu}^{(A^{\prime}
B^{\prime})} \epsilon^{AB},    \label{angular}
\end{equation}

\vskip10pt \noindent where circle brackets denote symmetrization,

\begin{equation}
\mu^{(AB)}:=\frac{1}{2}\epsilon_{A^{\prime}{B^{\prime}}} (M^{AA^{\prime
}BB^{\prime}}+M^{BA^{\prime}AB^{\prime}}),    \label{mu}
\end{equation}

\vskip10pt \noindent and where

\begin{equation}
\mu^{(AB)} \ = \zeta^{(A} \iota^{B)},  \label{general}
\end{equation}

\vskip10pt \noindent for some spinors $\zeta^{A}$ and $\iota^{B}$.

\vskip10pt \noindent From (\ref{Minkowskivector}) one notes also that a
simple spinor $\pi$, modulo multiplication by a phase factor, defines a
Lorentz four-vector $P$ with a Lorentz norm equal to zero, i.e.\@ a Lorentz
null four-vector (not just a null-direction)

\begin{equation}
P^{n}:= \pi^{N} {\bar\pi}^{N^{\prime}}.   \label{flagpole}
\end{equation}

\vskip10pt \noindent To obtain Lorentzian interpretation of the spinor phase
one observes then that a simple spinor also defines a unique (but
algebraically special) antisymetric Lorentz four-tensor of second rank

\begin{equation}
P^{ab} =-P^{ba}= P^{AA^{\prime}BB^{\prime}} = -P^{BB^{\prime}AA^{\prime}}=
\pi^{A} \pi^{B} \ \epsilon^{A^{\prime} B^{\prime} } + {\bar\pi}^{A^{\prime}} 
{\bar\pi}^{B^{\prime}} \epsilon^{AB}.    \label{flagplane}
\end{equation}

\vskip10pt \noindent$P^{a}$ in (\ref{flagpole}) is called the flagpole of a
(simple) spinor while $P^{ab}$ in (\ref{flagplane}) is known as its
flagplane. While a spinor defines its flagpole and its flagplane uniquely
(in a ``quadratic fashion''), from a given flagpole and its (algebraically
special) flagplane the corresponding spinor can be recovered only up to a
sign. This indicates again that the two dimensional spinors are more
elementary than Lorentz four-vectors.

\vskip10pt \noindent More details about how to manufacture Lorentz
four-tensors out of spinor-tensors using rules of the spinor algebra can be
found, for example, in Penrose's and Rindler's book \cite{prrw} (see also \cite
{Stewart} chapter 2).

\vskip10pt \noindent Spinors may also be introduced using the so called
Clifford algebra approach but the fundamental role of spinors in such an
approach is no longer as transparent as was outlined above. On the other
hand, the Clifford algebra approach may be easily generalised. It becomes
possible to construct spinors for real metric vector spaces of any dimension
and any non-degenerate signature\footnote{%
according to Professor Rafa{\l} Ab{\l}amowicz, the Clifford algebra and its
spinors may be introduced even when the metric is degenerate.}. For more
details about Clifford algebra constructions see, for example, the appendix in the
second volume of Penrose's and Rindler's book \cite{prrw}.

\section{TWISTORS, EVENTS AND PARTICLES.}
In this section the notion of the spinor space will be extended and
generalised in such a way so that it will become a four dimensional complex
vector space $\C^{4}$ equipped with a ``metric'' (a pseudo-hermitian form),
invariant with respect to the action of the $SU(2,2)$ matrix group. This
group contains a subgroup composed of elements of the form $L \oplus{\bar l}$%
, ${\bar l} \equiv{\bar L^{-1}}$ that belong to the matrix group in the
direct sum of $\mathrm{SL}(2,\C)$ and ${\overline{\mathrm{SL}(2,\C)}}$, 
i.e.\@ in $\mathrm{SL}(2,\C) \oplus {\overline{\mathrm{SL}(2,\C)}}$ 
(see (\ref{combine})). We will start from this
subgroup, which represents the covering of the identity component of the
Lorentz group, and successively fills up to the $SU(2,2)$ matrix
group. By using this fact we wish to show how the remaining parts of
$SU(2,2)$ correspond to four-translations, dilations and special conformal
transformation in the \emph{affine} Minkowski space. The complex vectors $%
\{ Z^{\alpha}, W^{\alpha},...\}$ of the complex vector space $\C^{4}$
equipped with $SU(2,2)$ invariant pseudo-hermitian form are called
(non-projective) twistors. Twistors are then geometrical vectors with
respect to the group $SU(2,2)$. To keep track of the physical interpretation
it is essential to represent a twistor (although not $SU(2,2)$ invariantly
but only ``Poincar{\'e}'' covariantly and ``Lorentz'' ($\mathrm{SL}(2,\C)$)
invariantly) in terms of two spinors. One consequence of this is that the
pseudo-hermitian form (of signature $++--$) preserved by $SU(2,2)$ will not
be diagonal in such a representation.

\vskip10pt \noindent First consider a direct sum of two spinor
spaces $T:=S \oplus{\bar S}^{*} \simeq \C^{4}$. From the analysis in the
previous section it follows directly that, geometrically, $T$ is invariant
with respect to the above mentioned subset of $\mathrm{SL}(2,\C) \oplus{\overline
{\mathrm{SL}(2,\C)}}$ matrix transformations. More explicitly, the action of the
``Lorentz'' group on an element in $S \oplus{\bar S}^{*}$ reads

\begin{equation}
\left( 
\begin{array}{c}
{\tilde\omega}^{B} \\ 
{\tilde\pi}_{B^{\prime}}
\end{array}
\right) = \left( 
\begin{array}{cc}
L^{B}_{A} & 0 \\ 
0 & {\bar l}^{A^{\prime}}_{B^{\prime}}
\end{array}
\right) \left( 
\begin{array}{c}
\omega^{A} \\ 
\pi_{A^{\prime}}
\end{array}
\right) ,  \label{bispinor}
\end{equation}

\vskip10pt \noindent where $Z^{\alpha}=\left( 
\begin{array}{c}
\omega^{A} \\ 
\pi_{A^{\prime}}
\end{array}
\right) $ represents a vector in $T$.

\vskip10pt \noindent The block diagonal transformations\footnote{%
with ${\bar l}={\bar L}^{-1}$.} in (\ref{bispinor}) preserve, of course, the
blockdiagonal ``metric''

\begin{equation}
\left( 
\begin{array}{cc}
\epsilon_{AB} & 0 \\ 
0 & {\bar\epsilon}^{{B^{\prime}}{A^{\prime}}}
\end{array}
\right)      \label{symp}
\end{equation}

\vskip10pt \noindent or, more explicitly,

\begin{equation}
\left( 
\begin{array}{cc}
L_{A}^{B} & 0 \\ 
0 & {\bar l}_{B^{\prime}}^{A^{\prime}}
\end{array}
\right) \left( 
\begin{array}{cc}
\epsilon_{BC} & 0 \\ 
0 & {\bar\epsilon}^{{C^{\prime}}{B^{\prime}}}
\end{array}
\right) \left( 
\begin{array}{cc}
L_{D}^{C} & 0 \\ 
0 & {\bar l}_{C^{\prime}}^{D^{\prime}}
\end{array}
\right) = \left( 
\begin{array}{cc}
\epsilon_{AD} & 0 \\ 
0 & {\bar\epsilon}^{{D^{\prime}}{A^{\prime}}}
\end{array}
\right) .          \label{symp1}
\end{equation}

\vskip10pt \noindent The elements in $T:=S \oplus{\bar S}^{*} \simeq \C^{4}$
acted upon by transformations of the form (\ref{bispinor}),
are geometrical objects with respect to the ``metric'' in (\ref{symp}), called 
\textbf{Dirac bispinors}. Resuming, we note that the
splitting of $T$ into a direct sum of two spinor spaces $S \oplus{\bar S}^{*}
$ is invariant with respect to the action of the (universal covering of the)
identity component of the homogeneous Lorentz group\footnote{%
how to represent the discrete symmetries of space-time, the charge
conjugation, space and/or time reflections is not discussed in this review,
certain aspects of these are however touched upon in \cite{jmp}.}.

\vskip10pt \noindent Now consider the action of the general linear group $%
GL(4, \C)$ on $\C^{4}$. Such an action does not preserve any ``metric'' but,
nevertheless, can be represented spinorially, if one so wishes, by

\begin{equation}
\left( 
\begin{array}{c}
{\tilde\omega}^{B} \\ 
{\tilde\pi}_{B^{\prime}}
\end{array}
\right) = \left( 
\begin{array}{cc}
R^{B}_{A} & Q^{A^{\prime}B} \\ 
U_{AB^{\prime}} & {K}^{A^{\prime}}_{B^{\prime}}
\end{array}
\right) \left( 
\begin{array}{c}
\omega^{A} \\ 
\pi_{A^{\prime}}
\end{array}
\right),            \label{L4C}
\end{equation}

\vskip10pt \noindent where there are no restrictions on the spinor-tensors $%
R, \ Q, \ U$ and $K$. Of course, the splitting of $\C^{4}$ into the two
spinor spaces $S \oplus{\bar S}^{*}$ is not invariant with the non-diagonal
blocks present in the $GL(4, \C)$ tansformations. Complex vectors in $\C^{4}$,
although (non-invariantly) represented by a pair of spinors should not
be confused with Dirac bispinors: they are \textbf{not Dirac bispinors}
anymore. Transformations such as in (\ref{L4C}) are just spinorial
representations of the general complex linear group $GL(4,\C)$ acting on $%
\C^{4}$. The two spinors just represent a complex vector $Z^{\alpha}$ in $%
\C^{4}$. The same abstract complex vector $Z^{\alpha}$ in $\C^{4}$ can be
represented by any pair of spinors related to each other by a transformation
such as in (\ref{L4C}). In this sense, every complex vector $Z^{\alpha}$ in $%
\C^{4}$ has two levels of representation: first, in terms of abstract spinors
and, second in terms of spinor coordinates with respect to a spin frame. It
is important to fully understand this point concerning the two levels of
representation of an arbitrary complex vector in $\C^{4}$. To a mathematician,
such a strange representation of vectors in $\C^{4}$ may seem artificial but
when the $GL(4,\C)$ gets restricted to its subgroup $SU(2,2)$, the
spinorially defined blocks in the $GL(4,\C)$ matrix in (\ref{L4C}) will
acquire a physical meaning, as we shall see below in $(\ref{combine})$ and $(%
\ref{conf})$. The complex vectors in $\C^{4}$ become geometrical objects with
respect to the ``metric'' (pseudo-hermitian form) preserved by $SU(2,2)$.

\vskip10pt \noindent We proceed to a presentation of $SU(2,2)$ matrix
transformations that constitutes a linear representation of the (four to one
covering of the) conformal symmetry group $C(1,3)$ of the (compactified)
affine Minkowski space\footnote{%
on the (compactification of the) affine $M$ the action of $C(1,3)$ is
non-linear.} $M$. In other words the conformal symmetry group of the
(compactified) affine Minkowski space appears as a four to one homomorphic
image of $SU(2,2)$. This homomorphism has been analysed at many occasions by
Roger Penrose and others, see especially \cite{jmp,prmc} and references
therein.

\vskip10pt \noindent The spinor representation of $SU(2,2)$ transformations
is essential for the blockwise identification of its physically important
subgroups such as (the covering of the identity connected element of) the
Lorentz subgroup, subgroup of translations (the composition of these two
forming the important (covering of the identity connected element of) the
Poincar{\'e} subgroup), the soubgroup of the dilations and the subgroup of
the special conformal transformations. All these subgroups live inside $%
SU(2,2)$ and act on its complex geometrical vectors $Z^{\alpha}$ in $\C^{4}$.
Using the spinor representation, we deduce that $SU(2,2)$ matrices $t^{\beta}_{\alpha}$ are
of the following form

\begin{equation}
t^{\beta}_{\alpha} := \left( 
\begin{array}{cc}
\delta^{B}_{D} \  & \ 0 \\ 
iC_{B^{\prime}D} \  & \ {\bar\delta}^{D^{\prime}}_{B^{\prime}}
\end{array}
\right) \left( 
\begin{array}{cc}
\ L^{D}_{E} \  & 0 \\ 
0 & {\bar l}^{E^{\prime}}_{D^{\prime}}
\end{array}
\right) \left( 
\begin{array}{cc}
d \ \delta^{E}_{M} \  & \ 0 \\ 
0 & \ \frac{1}{d} \ {\bar\delta}^{M^{\prime}}_{E^{\prime}}
\end{array}
\right) \left( 
\begin{array}{cc}
\delta^{M}_{A} \  & \ -iT^{A^{\prime}M} \\ 
0 & \ {\bar\delta}^{A^{\prime}}_{M^{\prime}}
\end{array}
\right) ,           \label{combine}
\end{equation}

\vskip10pt \noindent where in (\ref{combine}), the first matrix component
from the right corresponds to translations ($T^{a}=T^{A^{\prime}A}$) in the 
\emph{affine Minkowski space}, the second matrix component from the right
corresponds to scale changes ($d^{2}$, where $d$ is a non-zero real or
purely imaginary number) of all Lorentz vectors and all position
four-vectors in the \emph{affine Minkowski space}, the third matrix
component from the right represents (spinorial) Lorentz transformations ($%
L^{A}_{B}, \ {\bar l}^{B^{\prime}}_{A^{\prime}} $) and finally, the fourth
matrix component from the right corresponds to special conformal
transformations ($C_{A^{\prime}B}$) of the \emph{affine Minkowski space}.
Allowing all permutation of the four component matrices in (\ref{combine})
defines the entire set of $SU(2,2)$ matrices.

\vskip10pt \noindent A partial proof of this claim now follows.

\vskip10pt \noindent The first easy thing to note is that the
transformations resulting from compositions given by (\ref{combine})
(or given by any permutation of the four transformations in (\ref{combine})
building up the total transformation), do \emph{not preserve}
the metric in (\ref{symp}), in general. This happens because there are non-diagonal
spinor blocks in the representation of total matrix $t^{\beta}_{\alpha}$ in (%
\ref{combine}). The decomposition of $T$ into a direct sum of two spinor
spaces is, therefore, not invariant. What happens is that the two spinor
spaces mix with each other if translations and special conformal
transformations are performed. Consequently, as mentioned before, the
geometrical vectors (non-projective twistors) of $T$ (on which the total
transformation $t^{\beta}_{\alpha}$ of the type such as in (\ref{combine})
act) should not be confused with the Dirac bispinors. The second easy thing
to note is that the determinants of all total transformations $%
t^{\beta}_{\alpha}$, composed the four component matrices of the type as
in (\ref{combine}), are always equal to 1. This happens because
the determinat of any of its four matrix components is equal to 1, as
follows from a simple inspection of the representation of the component
matrices in (\ref{combine}). We will now show that the set of total
transformations of the type (\ref{combine}) is also unitary and
preserves the pseudo-hermitian form $g$ explicitly represented by\footnote{%
note that the signature of the matrices in (\ref{signature}) and (\ref
{signatureinverse}) is $++--$, easily verified by diagonalisation. The
spinor representation of vectors in $\C^{4}$ requires the representation of
the $SU(2,2)$ invariant pseudo-hermitian form (the ``metric'') to appear in
this non-diagonal spinorial disguise.}

\begin{equation}
g_{\alpha\beta}:= \left( 
\begin{array}{cc}
0 & {\bar\delta}_{A^{\prime}}^{B^{\prime}} \\ 
\delta^{A}_{B} & 0
\end{array}
\right)  ,      \label{signature}
\end{equation}

\vskip10pt \noindent and its inverse is given by

\begin{equation}
g^{\alpha\beta}:= \left( 
\begin{array}{cc}
0 & \delta^{B}_{A} \\ 
{\bar\delta}_{B^{\prime}}^{A^{\prime}} & 0
\end{array}
\right) .        \label{signatureinverse}
\end{equation}

\vskip10pt \noindent We have just to show that for all total transformations%
\footnote{%
any permutation of the the four matrix components in (\ref{combine}) gives
such a total transformation.} $t^{\beta}_{\alpha}$ of the type (%
\ref{combine}), the following properties are valid:

\begin{equation}
{\overline{{t}}}^{\beta}_{\delta} \ t^{\alpha}_{\gamma} \ \ g_{\alpha\beta}
= g_{\delta\gamma},   \label{twistinv}
\end{equation}

\begin{equation}
({{\overline{t}}^{T}})^{\alpha}_{\beta} \ = (t^{-1})^{\alpha}_{\beta} , \label{pseudoherm}
\end{equation}

\vskip10pt \noindent where the superscript index $T$ indicates the
transposition, the bar (as always in this paper) over a letter indicates
complex conjugation while the superscript $-1$ indicates the inverse.

%

\vskip10pt \noindent To verify identities in (\ref
{twistinv}) and (\ref{pseudoherm}), it is enough do it separately for each
component of the total transformation in (\ref{combine}). For
example, if we take the translation component of a transformation of the form (%
\ref{combine}), we verify (\ref{twistinv}) by noting that

\begin{equation}
\left( 
\begin{array}{cc}
{\bar\delta}^{A^{\prime}}_{B^{\prime}} & 0 \\ 
iT^{A^{\prime}B} & \delta^{B}_{A}
\end{array}
\right) \left( 
\begin{array}{cc}
0 & {\bar\delta}_{A^{\prime}}^{K^{\prime}} \\ 
\delta^{A}_{K} & 0
\end{array}
\right) \left( 
\begin{array}{cc}
\delta^{K}_{L} & -iT^{L}_{K^{\prime}} \\ 
0 & {\bar\delta}^{L^{\prime}}_{K^{\prime}}
\end{array}
\right) = \left( 
\begin{array}{cc}
0 & {\bar\delta}_{B^{\prime}}^{L^{\prime}} \\ 
\delta^{B}_{L} & 0
\end{array}
\right) .         \label{inva}
\end{equation}

\vskip10pt \noindent Identity (\ref{pseudoherm}) for the translation
component follows by a similar argument:

\begin{equation}
{\overline{\left( \left( 
\begin{array}{cc}
0 & {\bar\delta}_{A^{\prime}}^{K^{\prime}} \\ 
\delta^{A}_{K} & 0
\end{array}
\right) \left( 
\begin{array}{cc}
\delta^{K}_{M} & -iT^{M^{\prime}K} \\ 
0 & {\bar\delta}^{M^{\prime}}_{K^{\prime}}
\end{array}
\right) \left( 
\begin{array}{cc}
0 & \delta^{M}_{L} \\ 
{\bar\delta}_{M^{\prime}}^{L^{\prime}} & 0
\end{array}
\right) \right) }} ^{T} =        \label{inversetransl}
\end{equation}

\begin{equation*}
= {\overline{\left( 
\begin{array}{cc}
{\bar\delta}^{L^{\prime}}_{A^{\prime}} & 0 \\ 
-iT^{L^{\prime}A} & \delta^{L}_{A}
\end{array}
\right) }} ^{T} = \left( 
\begin{array}{cc}
\delta^{L}_{A} & iT^{A^{\prime}L} \\ 
0 & {\bar\delta}^{A^{\prime}}_{L^{\prime}}
\end{array}
\right) . 
\end{equation*}

\vskip10pt \noindent Here, the last matrix represents the inverse of
the translation, as asserted. The remaining verifications are left as a
spinor algebra exercise to the reader. This completes our partial proof of
the fact that the matrices arising as the result of multiplication of the
four components (taken in any order) displayed in (\ref{combine}) constitute
the matrix group $SU(2,2)$.

\vskip10pt \noindent For future reference, an explicit formula transforming
the spinor representatives of a twistor, and arising as a result of the
composition of the four symmetry transformations in (\ref{combine}), is
displayed here

\begin{equation*}
{\tilde Z}^{\alpha} = t^{\alpha}_{\beta}\ Z^{\beta} , 
\end{equation*}
or, equivalently,
\begin{equation}
\left( 
\begin{array}{c}
{\tilde\omega}^{B} \\ 
{\tilde\pi}_{B^{\prime}}
\end{array}
\right) = \left( 
\begin{array}{cc}
{d} L^{B}_{A} \  & \ - i {d} \ T^{A^{\prime}D} L^{B}_{D} \\ 
i {d} C_{B^{\prime}D}L^{D}_{A} \  & \ {d} \ C_{B^{\prime}D}
T^{A^{\prime}F}L^{D}_{F} + \frac{1}{d} \ {\bar l}^{A^{\prime}}_{B^{\prime}}
\end{array}
\right) \left( 
\begin{array}{c}
\omega^{A} \\ 
\pi_{A^{\prime}}
\end{array}
\right) .        \label{conf}
\end{equation}

\vskip10pt \noindent Consider now two contravariant twistors given by their
spinor representatives as follows

\begin{equation}
Z^{\alpha}:= \left( 
\begin{array}{c}
\omega^{A} \\ 
\pi_{A^{\prime}}
\end{array}
\right) , \ \ \ \ W^{\alpha}:= \left( 
\begin{array}{c}
\lambda^{A} \\ 
\eta_{A^{\prime}}
\end{array}
\right) .   \label{ZW}
\end{equation}

\vskip10pt \noindent Their ``covariant'' versions read

\begin{equation}
{\bar Z}_{\alpha}:= g_{\alpha\beta} {\overline{Z^{^{\beta}}}} = ({\bar\pi}%
_{A}, \ \ \ {\bar\omega}^{A^{\prime}}), \\
\ \ \ \ {\bar W}_{\alpha}:= g_{\alpha\beta} {\overline{Z^{^{\beta}}}} = ({%
\bar\eta}_{A}, \ \ {\bar\lambda}^{A^{\prime}}).     \label{ZWcovariant}
\end{equation}

\vskip10pt \noindent Using the spinor representation of the conformally ($%
SU(2,2)$) invariant complex valued ``scalar'' product (known by mathematicians\
as a pseudo-hermitian form) in the twistor space, we can write it as follows

\begin{equation}
\rho:=Z^{\alpha} \ {\bar W}_{\alpha} =g_{\alpha\beta} \ Z^{\alpha} \ {%
\overline{W^{^{\beta}}}} \ =\omega^{A} {\bar\eta}_{A} + {\bar\lambda }%
^{A^{\prime}} \pi_{A^{\prime}}.    \label{ZWscalar}
\end{equation}

\vskip10pt \noindent Note that the ``length'' (pseudo-hermitian norm) of any
twistor is always a real number, either positive or negative or null.

\vskip10pt \noindent As well-known to mathematicians, the imaginary part of
the pseudo-hermitian form preserved by the action of the matrix group $%
SU(2,2)$ defines on $T \simeq \C^{4}$ a conformally invariant symplectic
structure\footnote{%
simply meaning that conformal transformations (\i .e.\@$SU(2,2)$
transformations) form a closed subgroup of all canonical transformations
preserving such a symplectic stucture.} which may be expressed in terms of
global canonical and $SU(2,2)$ invariant Poisson bracket commutation
relations as

\begin{equation}
\{{\bar Z}_{\beta}, \ Z^{\alpha}\} = i{\delta}^{\alpha}_{\beta},  \label{PoissonZ}
\end{equation}

\vskip10pt \noindent with all the remaining Poisson bracket commutation
relations being equal to zero. In terms of the twistor's spinor
representatives it reads

\begin{equation}
\{{\bar\pi_{B}}, \ \omega^{A} \} = i{\delta}^{A}_{B}.   \label{Poissonomega}
\end{equation}

\vskip10pt \noindent The set of all twistors $\{Z, W, ...\}$ in $T$ modulo
multiplication by a non-zero complex number defines its projective
counterpart $\C P(3)$. Depending on the sign of the ``length''
(pseudo-hermitian norm) of a twistor, $\C P(3)$ is divided into three parts.
The first part of $\C P(3)$ corresponds to twistors of positive norm, the
second part to twistors of negative norm and the third to twistors with zero
norm.

\vskip10pt \noindent Up to now we have been quite silent about the physical
interpretation of the elements (complex vectors) in the twistor space $T
\simeq \C^{4}$ and their projective counterparts (complex lines through the
origin in $T$) in the projective twistor space $\C P(3)$, except for the fact
that they are natural carriers of the conformal (i.e.\@$SU(2,2)$) symmetry.
As shown by Penrose, each twistor may be interpreted in the Minkowski space
both geometrically and dynamically.

\vskip10pt \noindent Geometrically, any element in $\C P(3)$ can be identified
with a shear-free and, in general, twisted congruence (Robinson congruence)
of null-lines filling up the entire affine Minkowski space or, alternatively,
with the so called $\alpha$-plane in a complexified Minkowski space. In this\
paper, we will not discuss in detail the Minkowski geometrical interpetation\
of twistors, as we expect to obtain the Minkowski space time events (i.e.\@
affine \emph{position} four-vectors) themselves as twistor constructions. We
wish thus to follow the same pattern as in the previous section where
Lorentz four-vectors and, thereby, the entire Minkowski vector space appeared
as a subspace of the spinor-tensor algebra (namely, as a set of all hermitian
spinor-tensors of mixed second rank). More information on interpretation of
twistors as geometrical objects in (also curved) space-time can be found
in \cite{QG1,pr3,prmc,prrw}. In this context, the holomorphic
aspects of complex valued functions of twistors are very important, allowing
to perform geometrical constructions of general solutions to well-known
equations in physics, not to omit the famous non-linear (and anti-selfdual)
graviton construction. However, these fascinating topics, as said above,
will not be discussed in this paper.

\vskip10pt \noindent The dynamical/kinematical description of a twistor is
most easy to grasp by using the spinor representation of twistors and by
restricting (at least at the first instance) the $SU(2,2)$ matrix group
action to its ``Poincar{\'e}'' subgroup by putting $C_{A^{\prime}B}=0$ and $%
d=1$ in ($\ref{combine}$). We allow also the remaining two matrix components
of the matrix $t^{\alpha}_{\beta}$ in ($\ref{combine}$) to apear in the
reverse order. This defines completely (the covering group of the identity
connected part of) the Poincar{\'e} group. A very similar representation of
the Poin\-ca\-r\'e group was also considered by Bogoliubov, Todorov and
Lagunov in their monograph \cite{Todorov}.

\vskip10pt \noindent Using the spinor representatives of a twistor, it is
quite easy now to extract the dynamical/kinematical quantities. A twistor
defines the total four-angular momentum (including its orbital part and
therefore represented by a \emph{translation dependent} antisymmetric
Lorentz tensor of second rank) of a massless object, its total linear
four-momentum (a true Lorentz null four-vector) and also (which is a genuine
new feature) its (classical limit of the) helicity\footnote{%
note that the two dynamical quantities in (\ref{Poincare1})-(\ref{Poincare2}%
) are only Poincar{\'e} invariant (covariant) while the helicity is also a
conformally invariant scalar, as will be shown below in (\ref{helicity}).}:

\begin{equation}
P_{a} := \pi_{A^{\prime}}{\bar\pi}_{A} \ \ {\text{four-linear momentum}}, \label{Poincare1}
\end{equation}

\begin{equation}
M_{ab}:=i{\bar\omega}_{(A^{\prime}}\pi_{B^{\prime})}\epsilon_{AB}-i{\omega }%
_{(A}{\bar\pi}_{B^{\prime})}\epsilon_{A^{\prime} B^{\prime} } \ \ {\text{%
four-angular momentum}},   \label{Poincare2}    
\end{equation}

\vskip10pt \noindent where we used spinor representation of Lorentz tensors
in accordance with the previous section. Note that with respect to
translations represented by the first component from the right in (\ref
{combine}), the angular four momentum has correct transformation properties%
\footnote{%
exercise: show this using the Fierz identity in (\ref{Fierz}).}. The
(classical) helicity arises by extracting the translation invariant spin
contribution to the total angular four-momentum in (\ref{Poincare2}). This
is achieved in the usual way by forming the so called Pauli-Luba{\'n}ski
spin four-vector from (\ref{Poincare1})-(\ref{Poincare2}). By performing
some elementary spinor tensor algebra we obtain\footnote{%
try to fil in the details of this calculation yourself. Otherwise look it up
in \cite{prmc}.}

\begin{equation}
S^{a}:={\frac{1 }{2}}\epsilon^{abcd}M_{bc}P_{d}=sP^{a}=s\pi^{A^{\prime}}{%
\bar\pi}^{A} \ \ \text{where} \ \ s:= {\frac{1 }{2}}(Z^{\alpha}\bar
Z_{\alpha}).       \label{helicity}
\end{equation}

\vskip10pt \noindent A quite remarkable fact is that the real
valued conformally invariant function $s$ in (\ref{helicity}), i.e.\@ half of
the $SU(2,2)$ norm of $Z$, defines (the classical limit) of the helicity
operator of a massless object (particle).

\vskip10pt \noindent In addition, it turns out also that the canonical
conformally invariant Poisson bracket relations in (\ref{Poissonomega})
imply that the real valued, Poincar{\'e} covariant functions on $T$,
representing physical variables of a massless particle as spinorially
defined in (\ref{Poincare1})-(\ref{Poincare2}), fulfill the Poisson bracket
relations of the Poincar{\'e} algebra\footnote{%
a tedious spinor algebra calculation can prove that, you want to try? Do you
have another method to prove it?}:

\begin{equation}
\{P_{a},\ P_{b}\}=0,  \label{P4}
\end{equation}

\begin{equation}
\{M_{ab} ,\ P_{c}\}= -P_{a}{\eta}_{bc} + P_{b}{\eta}_{ac},  \label{MP4}
\end{equation}

\begin{equation}
\{M_{ab}, \ M_{cd}\}= M_{ac}\eta_{bd} + M_{bd}\eta_{ac} -M_{ad}\eta_{bc} -
M_{bc}\eta_{ad}.  \label{MM4}
\end{equation}

\vskip10pt \noindent It is obvious  that the \emph{non-invariant} translation spinor\
part, ``omega'', of a twistor has to do with the Minkowski position four-vector (take\
a look at the first matrix component from the right in (\ref{combine}) acting on\
twistors), which is the important element in the usual definition of the orbital part\
of the total angular four-momentum, here defined twistorially in (\ref{Poincare2}).\
To figure out where the Minkowski four-position variables may be hidden inside the
``omega'' part of a twistor we represent it spinorially as follows (by a
spinor-tensor contraction)\footnote{%
the imaginary number $i$ is inserted in order to harmonize with the the
adopted notational conventions as e.g.\@ the representation of the $SU(2,2)$
explicitly given in (\ref{combine}).}

\begin{equation}
\omega^{A}:=iz^{A{A^{\prime}}}\pi_{A^{\prime}}.  \label{omegaz}
\end{equation}

\vskip10pt \noindent In (\ref{omegaz}) $z^{A{A^{\prime}}}$ is a mixed spinor
tensor of second rank (not necessarily hermitian) that is equivalent with a
complexified Lorentz four-vector whose real part (i.e.\@ hermitian part of $%
z^{A{A^{\prime}}}$) transforms under the action of four translations (the
first matrix component from the right in (\ref{combine})) as a position
four-vector in the \emph{affine} Minkowski space. Therefore, the real part of 
$z$ is a good candidate to represent physical space-time events that
massless twistor objects trace out in the affine Minkowski space. However, $%
z^{A{A^{\prime}}}$ is not uniquely determined by the ``omega'' and ``pi''
parts of twistors because any complexified Lorentz four-vector (with its real part
defining a set of events in the affine Minkowski space) of the form

\begin{equation}
{\breve{z}}^{ AA^{\prime}}=z^{AA^{\prime}} + \lambda{\alpha}^{A}  \label{alphaplane}
\pi^{A^{\prime}},
\end{equation}

\vskip10pt \noindent where $\lambda$ is any complex valued parameter and
where ${\alpha}^{A}$ is an arbitrarily chosen spinor (zero spinor excluded),
would do.

\vskip10pt \noindent The equation in (\ref{alphaplane}) defines what is
nowadays called an $\alpha$-plane in the complexified Minkowski space. Each
``contravariant'' (with respect to the $SU(2,2)$ ``metric'') twistor $Z$
defines such a plane. The corresponding ``covariant'' (with respect to the $%
SU(2,2)$ ``metric'') twistor $\bar Z$ defines another such a plane that is
called $\beta$-plane.

\vskip10pt \noindent If the norm of a twistor vanishes (helicity in (\ref
{helicity}) equals zero), then complexified position spinor-tensors
(complexified position four-vectors) $z^{A{A^{\prime}}}$ and ${\breve z}^{A{%
A^{\prime}}}$ become hermitian (i.e.\@ complexified four-vectors become
real) and the two complex planes merge into a real line defining a single
null-line in the real affine Minkowski space:

\begin{equation}
{\breve x}^{AA^{\prime}}=x^{AA^{\prime}} + \lambda{\bar\pi}^{A} \pi
^{A^{\prime}},  \label{null-line}
\end{equation}

\vskip10pt \noindent where $\lambda$ is now an arbitrary \emph{real}
parameter. This is the famous correspondence between zero norm twistors and
null-lines in the Minkowski space \cite
{hl,MathSoc,oxf,pr2,jmp,prmc,prrw,perjes}. $\alpha$-~plane can also be
interpreted as the so called right handed twisting Robinson congruence in
the real Minkowski space while the corresponding $\beta$-~plane may then be
interpreted as almost the same twisting Robinson congruence but with the
twist reversed. Besides, the handedness of the twist is determined by the
sign of the twistor norm (remember this is just a real number, positive,
null or negative). As we mentioned above, we will not be very informative about
the space-time interpretations of twistors. Nevertheless, the above tiny piece of
information (without any proofs) may add some understanding to the
ideas involved in the twistor formalism.

\vskip10pt \noindent There exists a large number of articles about the
massless particles and massless fields described in terms of twistors, as
we briefly referred. The interested reader is adviced to go to
the original papers \cite{hl,MathSoc,oxf,pr2,jmp,prmc,prrw,perjes} and
references therein. We will, however, not dwell on this massless case much
more but turn to the case of the simplest (classical limit of the) massive
object that can be constructed in terms of twistors. The phase space nature
(symplectic structure) of the twistor space will be our main tool.

\vskip10pt \noindent For that reason consider now a direct product of
two-twistor spaces $T \times T$ with its diagonal deleted\footnote{%
in section two, in order to define a non-null Lorentz four-vector in terms
of spinors we needed (at least) two of them. The two were not allowed to be
proportional to each other. In the same way, to define a massive object in
terms of twistors we need (at least) two of them. They must not be
proportional to each other.}. The resulting space will be denoted by $T
\Delta T$. For each element $(Z,W)$ in $T \Delta T$ we require that $%
Z^{\alpha} \neq l \ W^{\alpha}$, where $l$ is any complex number. This can be\
formulated in another way: we require that the two twistors in the pair do not define
the same point in $\C P(3)$.

\vskip10pt \noindent  By linearity (see (\ref{PoissonZ})), the phase space structure\
on $T \Delta T$ is given by the following conformally
invariant global canonical Poisson bracket commutation relations

\begin{equation}
\{{\bar Z}_{\beta}, \ Z^{\alpha}\} = i{\delta}^{\alpha}_{\beta}, \qquad
\qquad\qquad\{{\bar W}_{\beta}, \ W^{\alpha}\} = i{\delta}^{\alpha}_{\beta} \label{PoissZW}
\end{equation}

\vskip10pt \noindent with all the remaining Poisson bracket commution
relations being equal to zero. In terms of the spinor representatives, the
two twistors are given by

\begin{equation}
Z^{\alpha} = (\omega^{A},\ \pi_{A^{\prime}}) \qquad\text{and} \qquad
W^{\alpha} = (\lambda^{A},\ \eta_{A^{\prime}}),  \label{ZWomegapilambdaeta}
\end{equation}

\vskip10pt \noindent so that the conformally invariant canonical global
Poisson bracket commutation relations in (\ref{PoissZW}) may be written (see
(\ref{Poissonomega})) as

\begin{equation}
\{{\bar\pi_{B}}, \ \omega^{A} \} = i{\delta}^{A}_{B}, \qquad\qquad \qquad\{{%
\bar\eta_{B}}, \ \lambda^{A} \} = i{\delta}^{A}_{B}   \label{inspinors}
\end{equation}

\vskip10pt \noindent with all the remaining Poisson bracket commution
relations being equal to zero.

\vskip10pt \noindent Two each twistor in the pair there corresponds an $%
\alpha$-plane (and also a $\beta$-plane) in the complexified Minkowski space.
Therefore, the intersection of the two non-coinciding, by definition because
the diagonal was excluded, $\alpha$-planes meet in a single complexified
$z$ position Lorentz four-vector. Therefore, an arbitrary point in $T \Delta T
$ defines such a $z$ explicitly. We just need to solve the following set of
equations

\begin{equation}
\omega^{A}=iz^{A{A^{\prime}}}\pi_{A^{\prime}}, \ \ \ \ \ \lambda ^{A}=iz^{A{%
A^{\prime}}}\eta_{A^{\prime}}.   \label{omegalambdaz}
\end{equation}

\vskip10pt \noindent The solution of the equations in (\ref{omegalambdaz})
reads (see \cite{prmc,prrw})

\begin{equation}
z^{AA^{\prime}}= {\frac{i}{f}}(\omega^{A}\eta^{A^{\prime}}-\lambda^{A}%
\pi^{A^{\prime}}),   \label{complexpos}
\end{equation}

\vskip10pt \noindent where 
\begin{equation}
f: = \pi^{A^{\prime}}\eta_{A^{\prime}} \neq0,   \label{f}
\end{equation}

\vskip10pt \noindent the last inequality being true just because we are in $%
T \Delta T$.

\vskip10pt \noindent It is tempting to define the Lorentz position
four-vector as the hermitian part of the solution (i.e.\@ the real part of
the corresponding complexified position four-vector) in (\ref{complexpos}).
This would mean that we take the complex intersection point of the two $%
\alpha$ planes defined by the two ``contravariant'' twistors $Z, W$ and the
complex intersection of the corresponding two $\beta$ planes defined by the
corresponding ``covariant'' twistors ${\bar Z}, \ {\bar W}$ and take the
mean value of them which is then hermitian, i.e.\@ defines a real position
four-vector. Explicitly, such a Poincar{\'e} covariant hermitian (i.e.\@ real
position four-vector) spinor-tensor of mixed second rank $q$ reads

\begin{equation}
q^{a}=q^{AA^{\prime}}:={\frac{1}{2}}(z^{AA^{\prime}} + {\bar z}^{A^{\prime}
A} ) ,      \label{4poswithspin}
\end{equation}

\vskip10pt \noindent where, of course,

\begin{equation}
{\bar z}^{A^{\prime} A} = {\frac{-i}{{\bar f}}}({\bar\omega}^{A^{\prime}}{%
\bar\eta}^{A}- {\bar\lambda}^{A^{\prime}}{\bar\pi}^{A} ).  \label{complexconjugatepos}
\end{equation}

\vskip10pt \noindent Before we discuss further the relevance of the
definition of the position four-vector $q$ in (\ref{4poswithspin}) we
identify the angular four-momentum and linear four-momentum of the
two-twistor object in $T \Delta T$. This must be done in such a way that
the symplectic structure on $T \Delta T$ inherited from the symplectic
structure on $T$ (being the imaginary part of the pseudo-hermitian metric on 
$T$, preserved by $SU(2,2)$) still implies that they fulfill the commutation
relations of the Poincar{\'e} Poisson bracket algebra as in (\ref{P4}) - (%
\ref{MM4}). The task is easy because, by linearity, we can simply add
variables describing the two (mutually Poisson commuting) massless parts
(see (\ref{Poincare1})-(\ref{Poincare2})) and obtain
\begin{equation}
P_{a}=P_{A^{\prime}A} := \pi_{A^{\prime}}{\bar\pi}_{A} + \eta_{A^{\prime}}{%
\bar\eta}_{A} ,  \label{Poincare1+2P}
\end{equation}
\begin{eqnarray}
M_{ab} & = & M_{A^{\prime}A B^{\prime}B}  \nonumber \\
 & :=  & i({\bar\omega}_{(A^{\prime}}\pi
_{B^{\prime})}\epsilon_{AB}- {\omega}_{(A}{\bar\pi}_{B)}\epsilon_{A^{\prime}
B^{\prime} } )+i({\bar\lambda}_{(A^{\prime}}\eta_{B^{\prime})}\epsilon _{AB}-%
{\lambda}_{(A}{\bar\eta}_{B)}\epsilon_{A^{\prime} B^{\prime} } ). \label{Poincare1+2M}
\end{eqnarray}

\vskip10pt \noindent The angular and linear four-momenta defined in (\ref
{Poincare1+2P}) - (\ref{Poincare1+2M}) will automatically fulfil the
commution relations of the Poincar{\'e} Poisson bracket algebra. Using our
definition of the position four-vector in (\ref{4poswithspin}), it turns out
that the angular four momentum in (\ref{Poincare1+2M}) can be written as%
\footnote{%
to verify this requires a tedious spinor algebra calculation.}

\begin{equation}
M^{ab}=P^{a} \ q^{b} - P^{b} \ q^{a} + S^{ab},  \label{orbital+Lubanski}
\end{equation}

\begin{equation*}
S^{ab}:={\frac{1 }{{(P^{k}P_{k})}}}\epsilon^{abcd}P_{c}S_{d}, 
\end{equation*}

\vskip10pt \noindent and where the Pauli-Luba{\'n}ski four-vector $S^{a}$ 
is defined in the usual way,

\begin{equation}
S^{a}:={\frac{1 }{2}}\epsilon^{abcd}M_{bc}P_{d}.  \label{Pauli-Lubanski}
\end{equation}

\vskip10pt \noindent The fact displayed in (\ref{orbital+Lubanski}) would
support our choice of the definition of the position four-vector variable in
the phase space $T \Delta T$ according to (\ref{4poswithspin}). However,
there is one great disadvantage because the so defined position four-vector
is non-commuting; instead, the following Poisson bracket commutation
relations may be derived \cite{zab,zab1,zak}:

\begin{equation}
\{q^{a}, \ q^{b}\}=-S^{ab}.   \label{non-commutq}
\end{equation}

\vskip10pt \noindent The above non-commuting feature of the four-position
variable, in the case when the intrinsic angular momentum part is assumed to be
effectively defined by $S^{a}$ representing only three variables ($S^{a}$ in
(\ref{Pauli-Lubanski}) automatically fulfils $S^{a}P_{a}=0$), seems to be an
entirely generic feature in the geometry of the so called relativistic
extended phase spaces as discovered by professor Stanis{\l}aw Zakrzewski 
\cite{zak,zak2,zak1}. His discoveries (being an extension and generalisation
of certain mathematical results obtained by J.M. Souriau \cite{Sou}) were
made without any use of twistors and classify all Poincar{\'e} invariant
extended phase spaces. It has been proved in \cite{zab,zab1} that one of the
cases considered by him is a ``subset'' of the two-twistor construction as
presented in this report.

\vskip10pt \noindent Our conjecture is that with three or more twistors the
Zakrzewski's general feature would reappear in the twistor formalism and
perhaps could shed some light on the physical relevance of Zakrzewski's and
Souriau's mathematical achievements.

\vskip10pt \noindent The four-position $q$, in (\ref{4poswithspin}%
), will be called the centre of mass of the massive spinning and charged
system defined by a point in $T \Delta T$. It has been discovered in \cite
{zab,zab1} that a redefinition of the position four-vector may be found in
such way that the commutation of the four-position variables will be
restored. Defining a new four-position:

\vskip10pt \noindent
\begin{equation}
x^{a}= x^{AA^{\prime}}:= q^{AA^{\prime} } + \Delta x^{AA^{\prime}},  \label{4pos}
\end{equation}

\vskip10pt \noindent where

\begin{equation}
\Delta x^{AA^{\prime}}:= {\frac{i }{2f {\bar f}}}(\rho{\bar\pi}^{A}
\eta^{A^{\prime}} - {\bar\rho} \pi^{A^{\prime}}{\bar\eta}^{A}),  \label{Deltapos}
\end{equation}

\vskip10pt \noindent we obtain \cite{zab,zab1} that\footnote{%
all these statements may be checked by hand using the spinor representatives
of the twistors and the spinor algebra rules but this is tedious and not
very amusing. The ambitious reader is encouraged to do the necessary
calculations and find possible minor omissions or/and sign errors if any.
There is also the so called Penrose's blob notation that simplifies such
calculations a lot but first you have to be able to master the blob notation
:).}

\begin{equation}
\{x^{a}, \ x^{b}\}=0.  \label{commutx}
\end{equation}

\vskip10pt \noindent The new commuting four-position $x$ (in (\ref{4pos}))
shifted by $\Delta x$ (in (\ref{Deltapos})) away from the (non-commuting)
centre of mass $q$ (in (\ref{4poswithspin})) will be called the centre of
charge of the system. The total angular four-momentum now splits in a
different way compared with the splitting in (\ref{orbital+Lubanski})
because the orbital momentum is defined with respect to the centre of charge
instead. We obtain \cite{zab,zab1}:

\begin{equation}
M^{ab}=P^{a} x^{b} - P^{b} x^{a} + \Sigma^{ab},  \label{orbital-Dirac}
\end{equation}

\vskip10pt 
\noindent 
where 

\begin{equation}\label{Sigma}
\Sigma_{ab}:= [{\sigma}_{(A^{\prime}}\eta_{B^{\prime})}
\epsilon_{AB} + {\bar \sigma}_{(A} {\bar \eta}_{B)} 
{\bar \epsilon}_{A^{\prime}B^{\prime}}
 ],  
\end{equation}

\begin{equation}
{\sigma}_{A^{\prime}}:={\frac{i }{f}}(k \pi_{A^{\prime}}+\rho\eta
_{A^{\prime}}),  \label{sigm}
\end{equation}

\begin{equation}
k:=s_{1}-s_{2},   \label{k}
\end{equation}

\begin{equation}
s_{1} := {\frac{1 }{2}}(Z^{\alpha}\bar Z_{\alpha}), \qquad\qquad s_{2} := {%
\frac{1 }{2}}(W^{\alpha}\bar W_{\alpha}), \qquad\qquad\rho:=(Z^{\alpha}\bar
W_{\alpha}).  \label{2s}
\end{equation}

\vskip10pt \noindent It is relatively easy to realize that the translation
invariant antisymmetric spin four-tensor $\Sigma$ is composed of two parts:
one part defining rotation of the centre of mass around the centre of charge%
\footnote{%
this could be the root of the so called ``Zitterbewegung'', see our further
developments in this review and \cite{Corben}.}, and one part defining the
intrinsic rotation of the centre of charge itself. The number of variables
it defines is five and not six, as one would expect from a general
antisymmetric Lorentz four-tensor of rank two (but even five is contrasting
with the three variables defined by the Pauli-Luba{\'n}ski Lorentz
four-vector $S^{a}$ in (\ref{Pauli-Lubanski})) because of the vanishing of
the Poincar{\'e} scalar\footnote{%
this feature is a consequence of the two-twistor construction, with three or
more twistors, perhaps the invariant in (\ref{dualsigma}) would be different
from zero.}

\begin{equation}
\Sigma^{ab}\epsilon_{abcd} \Sigma^{cd} =0,   \label{dualsigma}
\end{equation}

\vskip10pt \noindent which may also be verified by direct spinor algebra
calculations. Note also that

\begin{equation}
\Sigma^{ab} \Sigma_{ab} = - 2 [(\sigma^{A^{\prime}}
\eta_{A^{\prime}})^{2} +({\bar\sigma}^{A} {\bar\eta}_{A})^{2}] = 4 k^{2},  \label{absvalue}
\end{equation}

\vskip10pt \noindent where, remarkably, $\Sigma^{ab} \Sigma_{ab} = 4k^{2}$ is
not only Poincar{\'e} but also conformally invariant scalar function.

\vskip10pt \noindent The variables $x$ in (\ref{4pos}) and $P$ in (\ref{Poincare1+2P})
define the eight dimensional phase space that can be identified with
the cotangent bundle of the real Minkowski space. The implied (by (\ref
{inspinors})) Poisson bracket commutation relations read

\begin{equation}
\{P_{a}, \ x^{b} \}= \delta^{b}_{a}, \label{Px}
\end{equation}

\vskip10pt \noindent as they should. Moreover, $\Sigma$ commutes with $x$ in (\ref
{4pos}) and with $P$ in (\ref{Poincare1+2P}) as shown in detail in \cite
{zab,zab1}. $\Sigma$ defines a second rank antisymmetric Lorentz four-tensor
valued function on a six dimensional Poincar{\'e} invariant phase space
identified with the cotangent bundle of the real projective
spinor space \cite{zab}, the latter spanned by the Poincar{\'e} invariant spinor
variable $\eta$ (of the second twistor $W$ in the pair) modulo its
multiplication by a non-zero real number. The Poincar{\'e} invariant
variable in the cotangent fiber is then the above defined $\sigma$ spinor
(in ({\ref{sigm})) modulo its multiplication by the inverse of this real
number. $\Sigma$ fulfills the commutation relations of the Lorentz Poisson
bracket algebra,

\begin{equation}
\{\Sigma_{ab}, \ \Sigma_{cd}\}= \Sigma_{ac}\eta_{bd} + \Sigma_{bd}\eta_{ac}
-\Sigma_{ad}\eta_{bc} - \Sigma_{bc}\eta_{ad}.  \label{Sigmacommute}
\end{equation}

\vskip10pt \noindent as it should do. The commutations relations in (\ref
{Sigmacommute}) can be computed by using the canonical commutation
relations defining the Poincar{\'e} invariant symplectic structure on the
cotangent bundle of the real spinor space (spinor modulo its multiplication
by a real number) or more simply using (\ref{PoissZW}) alternatively (\ref{inspinors}). 

\vskip10pt \noindent Finally, the conformally invariant scalar function

\begin{equation}
e:=2s_{1},      \label{e}
\end{equation}

\vskip10pt \noindent (we call it the classical limit of the electric charge quantum\
operator\footnote{%
remarkably, it appears as a conformal scalar, twice the helicity of the first
massless constituent, i.e.\@ as the norm of the first twistor in the pair.
Here our identification of the charge differs from that usually assumed by
people in the Penrose's group, see, for example, \cite{hl,perjes,sparling,tod1}.
They assume $s_{1}+s_{2}$ to be the electric charge but such a choice would
destroy our symplectic Poincar{\'e} invariant decomposition of $T \Delta T$.}
with $s_{1}$ defined in (\ref{2s})) and $\arg f$ (where $f$, the Poincar{\'e}
invariant scalar function, was defined in (\ref{f})), commute with $x$ in (%
\ref{4pos}) and $P$ in (\ref{Poincare1+2P}) and with all the variables of
the cotangent bundle of the real spinor space, as defined above, so that they
form an independent cotangent bundle over a circle . The implied commutation
relations read

\begin{equation}
\{e, \ {\arg f} \}=1.  \label{charge}
\end{equation}

\vskip10pt \noindent We have in this way (details in \cite{zab,zab1})
constructed a Poincar{\'e} invariant decomposition of the two twistor $T
\Delta T$ phase space (16 dimensions) into three independent parts: a
cotangent bundle of the Minkowski space (8 dimensions), a cotangent bundle
of a real projective spinor space (6 dimensions) and a cotangent bundle over
a circle (2 dimensions). The first two parts forming a 14 dimensional
extended phase space, i.e.\@ the cotangent bundle of the Minkowski space (8
dimensions) and the cotangent bundle of a real projective spinor space (6
dimensions) constitute (as already mentioned above) a special case of the
very general geometrical mathematical construction obtained, without any use
of twistors, by professor Stanis{\l}aw Zakrzewski in \cite{zak,zak2,zak1}.

\vskip10pt \noindent There is a peculiar discrete symmetry in our
construction. The entire construction works equally well if we change the
sign in front of the shift $\Delta x^{a}$ in (\ref{4pos}) and interchange
simulantaneously the role of the two spinors $\pi\rightarrow\eta$, $%
\eta\rightarrow\pi$. The charge variable will then be defined by the norm of
the second twistor (that is, $e:=2s_{2}$) instead. Further analysis of this
peculiarity and its possible interpretation in particle physics could
perhaps be of value.

\vskip10pt \noindent The decomposition of the phase space $T \Delta T$ as
constructed above\footnote{%
in much greater detail described in \cite{zab,zab1}.} is Poincar{\'e}
invariant. However, it is not invariant with respect to the special conformal
transformations, i.e.\@ it is not invariant with respect to the action of
the fourth component matrix from the right in (\ref{combine}). Only the
value of the charge $e$, the value of $\Sigma^{ab}\Sigma_{ab}=4k^{2}$ and, in
fact, the Lorentz norm of the Pauli-Luba{\'n}ski spin four-vector divided by
the norm of the (canonical) linear momentum four-vector $P^{a}P_{a}=2f{\bar f%
}$:

\begin{equation}
w^{2}:={\frac{{-S_{a}S^{a}}}{{P^{b}P_{b}}}},   \label{w1}
\end{equation}

\vskip10pt \noindent are, by definition, also preserved by the action of the
special conformal transformations (and dilations). The proof of the
conformal invariance of $w^{2}$ in (\ref{w1}) follows from a straight
forward spinor algebra calculations. We get that \cite{perjes,ab3}

\begin{equation}
w^{2} = {k^{2} + \rho{\bar\rho}},   \label{w2}
\end{equation}

\vskip10pt \noindent where $k$ and $\rho$ were previously defined in (\ref{k}%
) and in (\ref{ZWscalar}). They are, by definition, conformal ($SU(2,2)$)
scalar invariants.

\vskip10pt \noindent It is therefore interesting to see how the action of
the fourth component matrix from the right in (\ref{combine}) affect the
Poincar{\'e} covariant/invariant variables such as $q$, $x$, $\Delta x$, $P$%
, $S$ and $\Sigma$ defined in (\ref{4poswithspin}), (\ref{4pos}), (\ref
{Deltapos}), (\ref{Poincare1+2P}), (\ref{Pauli-Lubanski}), (\ref{Sigma}).

\vskip10pt \noindent For that reason we will determine how the conformal
transformations affect $z$ (i.e. the intersection of the two $\alpha$ planes
defined by the two twistors) and how they affect the Poincar{\'e}
invariant/covariant\footnote{%
we use the word invariant and covariant interchangebly; if by a
transformation we mean a change of the coordinate system then the
geometrical objects are invariant, only their coordinates are changing; on
the other hand, if the coordinate system is kept unchanged then every
geometrical objects is covariantly transformed into another one; the two
meanings merge into one when it comes to geometrical scalars.} spinors $\pi$%
, $\eta$. Some lengthy spinor algebra calculations give

\begin{equation}
{\tilde z}^{a}=\frac{z^{a}-\frac{1}{2}C^{a}(z_{k}z^{k})}{1-C_{b}z^{b} + 
\frac{1}{4}(C_{n}C^{n})(z_{m}z^{m}) },    \label{zconformal}
\end{equation}

\vskip10pt \noindent where we used the identifications: $z^{a}=z^{AA^{%
\prime}}$ (see (\ref{complexpos})) and $C_{a}=C_{A^{\prime}A}$ (see (\ref
{combine})). Note that $C$ is a real Lorentz four-vector while $z$, in
general, is not, with its real part (in the affine Minkowski space)
representing the (non-commuting) position four-vector $q$ of the centre of
mass of the two-twistor particle (spinning and electrically charged object)%
\footnote{%
it should perhaps be mentioned that the coordinate function of the
complexified position four-vector in (\ref{complexpos}) do, in fact, commute,
which is easy to see from (\ref{inspinors}).}.

\vskip10pt \noindent The special conformal transformation of the
intersection of the corresponding two $\beta$ planes, defined by the two
twistors, is obtained by a simple complex conjugation of (\ref{zconformal}).
If the norms of the two twistors $s_{1}$, $s_{2}$ and the ``scalar'' product 
$\rho$ all vanish, i.e.\@ if $s_{1}=s_{2}=\rho=0$, then the two-twistor
particle is still massive but uncharged and non-spinning. Its $z$ Lorentz
four-vector becomes real \cite{prmc}, i.e.\@ the two $\alpha$ and the two $%
\beta$ planes merge into two distinct null-lines with a real intersection
defining the Lorentz four-position $z=q=x$ transforming under special
conformal transformations in the well-known traditional way.

\vskip10pt \noindent For an arbitrary element in $T \Delta T$, i.e.\@ for a
general relativistic massive, spinning and charged object, however, our
definition of the Minkowski position four-vectors goes beyond the usual
definition because under the action of special conformal transformations
(the fourth matrix component in the righthandside of (\ref{combine})), the real
Lorentz position four-vector $q$ mixes with the spin, electric charge and
the linear momentum variables of the two-twistor particle in the way
indicated by (\ref{zconformal}). To see that one should recall that the
imaginary part of $z$ contains variables describing spin and charge and is
given by \cite{ab3,zab,zab1,hl,tod1}:

\begin{equation}
Y^{a} : = {\frac{1}{2i}}(z^{a} - {\bar z}^{a}) ={\frac{1}{2f\bar f}} [ \rho{%
\bar w}^{a} + {\bar\rho}w^{a} - 2s_{1}(P_{2})^{a} - 2s_{2}(P_{1})^{a} ] ,  \label{imaginaryz}
\end{equation}

\vskip10pt \noindent where

\begin{equation}
(P_{1})_{a} := \pi_{A^{\prime}}{\bar\pi}_{A}, \ \ \ \ (P_{2})_{a} :=
\eta_{A^{\prime}}{\bar\eta}_{A}, \ \ \ \ w_{a} := \pi_{A^{\prime}}{\bar\eta }%
_{A}.             \label{null}
\end{equation}

\vskip10pt \noindent The action of a special conformal transformastion on
the Poincar{\'e} invariant/covariant spinors $\pi$ and $\eta$ can be
expressed explicitly as

\begin{equation}
{\tilde\pi}_{A^{\prime}}= \pi_{A^{\prime}} + iC_{A^{\prime}A} \omega^{A}= [1-%
\frac{1}{2} C_{a}z^{a} ] \ \pi_{A^{\prime}} + \frac{1}{2} {\bar\epsilon }%
^{AB}(C_{AA^{\prime}}z_{BB^{\prime}} -C_{BB^{\prime}}z_{AA^{\prime}}) \
\pi^{B^{\prime}},  \label{piconformaltransformation}
\end{equation}

\begin{equation}
{\tilde\eta}_{A^{\prime}}= \pi_{A^{\prime}} + iC_{A^{\prime}A} \omega^{A}=
[1-\frac{1}{2} C_{a}z^{a} ] \ \eta_{A^{\prime}} + \frac{1}{2} {\bar\epsilon }%
^{AB}(C_{AA^{\prime}}z_{BB^{\prime}} -C_{BB^{\prime}}z_{AA^{\prime}}) \
\eta^{B^{\prime}}.   \label{etaconformaltransformation}
\end{equation}

\vskip10pt \noindent These formulae 
follow directly from (\ref{combine}).

\vskip10pt \noindent The transformation properties of the two spinors%
\footnote{%
in (\ref{Poincare1+2P}) they were used to define the linear four-momentum,
i.e.\@ it represented an element in the ``cotangent'' fiber over the
four-position (in the base manifold) of the massive two-twistor relativistic
object. The special conformal transformations destroy our Poincar{\'e}
invariant symplectic decomposition of $T \Delta T$ mapping it onto a new one.%
} $\pi$, $\eta$ under the action of the special conformal transformations ((%
\ref{piconformaltransformation}), (\ref{etaconformaltransformation})) are
new features arising as a consequence of the suggested twistor
representation of (the classical limit of) the relativistic physical
variables. We think that more conclusions could be drawn from this
observation. However, we will not investigate the issue any further in this
review.

\vskip10pt \noindent We proceed to an application of the formalism developed
above. In the next two sections, the Poincar{\'e} invariant dynamics of a
relativistic massive, charged and spining two-twistor particle will be
formulated on $T \Delta T$.

\section{MINIMALLY COUPLED SECOND ORDER DIRAC OPERATOR AND\
ITS CLASICAL LIMIT ON THE TWO-TWISTOR SPACE.}

\vskip10pt \noindent First we present some well-known facts (see e.g.\@\cite
{Schiff} and \cite{Stewart} pp.\@88-89 and Appendix A) about the
minimally coupled Dirac equation using the language of spinors developed in
the second section of the paper. The electro-magnetic Lorentz four-vector
potential expressed as an hermitian second rank mixed tensor field reads

\begin{equation}
A_{c}(x^{d})=A_{CC^{\prime}}( x^{DD^{\prime}} ),  \label{A}
\end{equation}

\vskip10pt \noindent where $x^{DD^{\prime}}$ is a translation dependent
hermitian mixed second rank spinor-tensor representing position four-vectors,
i.e.\@ labeling points in the Minkowski affine space. At this stage we do
not need to associate them with twistors. The components of $x^{DD^{\prime}}$
are then mutually commuting by definition.

\vskip10pt 
\noindent 
In relativistic quantum mechanics the (mechanical and
not ca{\-}nonical) linear four-momen{\-}tum operator of a charged massive quantum
particle in an external electro-magnetic field is defined by\footnote{%
we choose units so that $c= \hbar= 1$.}

\begin{equation}
{\hat p}_{BA^{\prime}}: =i\frac{\partial}{\partial x^{BA^{\prime}}} - e
A_{BA^{\prime}} (x^{CD^{\prime}}),  \label{mechpop}
\end{equation}

\vskip10pt \noindent where $e$ represents the numerical value of the
particle's electric charge. The canonical differential operator defined by

\begin{equation}
i\partial_{KM^{\prime}}:=i\frac{\partial}{\partial x^{KM^{\prime}}},  \label{canonpop}
\end{equation}

\vskip10pt \noindent represents particle's canonical linear four-momentum
operator (being essentially the same as the infinitesimal generator of
space-time four-translations) in the absence of interactions. The definition
in (\ref{mechpop}), on the other hand, represents the \emph{\textbf{minimal
coupling}} of the charged particle\footnote{%
with the value of its charge being equal to $e$.} to an external
electro-magnetic field defined by its potential four-vector $A_{a}(x^{b})$.

\vskip10pt \noindent A Dirac bispinor field\footnote{%
at this first quantized stage the components of the spinors are not
anticommuting operator valued distributions but usual complex number fields.}
over the Minkowski space

\begin{equation}
\Psi^{A}:= \Psi^{A}(x^{CD^{\prime}}) , \ \ \ \ \ \ \
\Phi_{A^{\prime}}:=\Phi_{A^{\prime}}(x^{CD^{\prime}})   \label{Diracbispinor}
\end{equation}

\vskip10pt \noindent is said to obey the Dirac equation if

\begin{equation}
{\hat p}^{AB^{\prime}} \ \Phi_{B^{\prime}} = \frac{m}{\sqrt{2}} \ {\Psi}%
^{A}, \ \ \ \ \ \ \ {\hat p}_{BA^{\prime}} \ {\Psi}^{B} = \frac{m}{\sqrt{2}}
\ \Phi_{A^{\prime}}.        \label{Dirac}
\end{equation}  

\vskip10pt \noindent Here (\ref{Dirac}) is interpreted as a classical (i.e.\@
non-quantum) formula and it the bispinors had constant values, then one
could say, with reference to the discussion in section two of this review,
that the Dirac equation simply states that the linear four-momentum is a sum
of two null vectors constructed from two distinct spinors $\Phi$ and $\Psi$
with their $\mathrm{SL}(2,\C)$ ``scalar'' product normalised to a real value\footnote{$%
\Psi ^{A}{\bar\Phi}_{A}=\frac{m}{\sqrt{2}}$.} $\frac{m}{\sqrt{2}}$ where $m$
is identified with the mass of the particle (this interpretation would
exclude the Majorana-Weyl bispinor because in this case the $\mathrm{SL}(2,\C)$
``scalar'' product is equal to zero). The Dirac equation can be viewed also as
an eigenvalue problem involving the mass $m$ as the spectral parameter because it\
can be written in a suggestive way as

\begin{equation}
\left( 
\begin{array}{cc}
0 & {\hat p}^{AB^{\prime}} \\ 
{\hat p}_{BA^{\prime}} & 0
\end{array}
\right) \left( 
\begin{array}{c}
\Psi^{B} \\ 
\Phi_{B^{\prime}}
\end{array}
\right) = \frac{m}{\sqrt{2}} \left( 
\begin{array}{c}
\Psi^{A} \\ 
\Phi_{A^{\prime}}
\end{array}
\right) .   \label{Diraceigen}
\end{equation}

\vskip10pt \noindent Looked upon in this way (i.e.\@ as in (\ref{Diraceigen}%
)), the linear four-momentum operator acts as an infinitesimal
four-translation operator on the components of the bispinor fields,
simultaneously mixing the two spinor fields with each other. In the so
called second order formulation of the Dirac equation the mixing can be
avoided and the Dirac equation gets a clear structure of an eigen-value
equation for the mass (squared)\footnote{%
which then constitutes a generalisation of the Klein-Gordon equation. The
latter describes a non-spinning charged (scalar) relativistic particle in an
external electro-magnetic field. Note also that the classical limit of the
Klein-Gordon equation reproduces the Lorentz force equation as will be
discussed shortly.}. The second order formulation of the Dirac equation is
easily obtained from (\ref{Dirac}), (\ref{Diraceigen}) and formally we get

\begin{equation}
{\hat p}_{BA^{\prime}} \ {\hat p}^{BB^{\prime}} \ \Phi_{B^{\prime}} = \frac{%
m^{2}}{2} \ \Phi_{A^{\prime}}, \ \ \ \ \ \ \ {\hat p}^{AB^{\prime}} \ {\hat p%
}_{BB^{\prime}} \ {\Psi}^{B} = \frac{m^{2}}{2} \ {\Psi}^{A},  \label{Dirac2}
\end{equation}

\vskip10pt \noindent which is equivalent to

\begin{equation}
\left( 
\begin{array}{cc}
0 & {\hat p}^{CA^{\prime}} \\ 
{\hat p}_{AC^{\prime}} & 0
\end{array}
\right) \left( 
\begin{array}{cc}
0 & {\hat p}^{AB^{\prime}} \\ 
{\hat p}_{BA^{\prime}} & 0
\end{array}
\right) \left( 
\begin{array}{c}
\Psi^{B} \\ 
\Phi_{B^{\prime}}
\end{array}
\right) =\frac{m^{2}}{2} \left( 
\begin{array}{c}
\Psi^{C} \\ 
\Phi_{C^{\prime}}
\end{array}
\right) .    \label{Dirac3}
\end{equation}

\vskip10pt \noindent We wish to obtain a more physical interpretation of 
Dirac equation (\ref{Dirac2}) (or, equivalently, that in (\ref{Dirac3}) ).
Therefore, we rewrite it as follows

\begin{eqnarray}
{\hat p}_{BA^{\prime}} \ {\hat p}^{BB^{\prime}} \Phi_{B^{\prime}} & =  & [ \frac
{1}{2} ( {\hat p}_{BA^{\prime}} \ {\hat p}^{BB^{\prime}} \ - {\hat p}_{B}^{\
B^{\prime}} \ {\hat p}^{B}_{\ A^{\prime}} ) + \frac{1}{2} ({\hat p}%
_{BA^{\prime}} \ {\hat p}^{BB^{\prime}} + {\hat p}_{B}^{\ B^{\prime}} \ {%
\hat p}^{B}_{\ A^{\prime}} )] \Phi_{B^{\prime}}   \nonumber \\  
 & = & \frac{m^{2}}{2} \ \Phi_{A^{\prime}},    \label{Dirac4}
\end{eqnarray}

\begin{eqnarray}
{\hat p}^{AB^{\prime}} \ {\hat p}_{BB^{\prime}} \ {\Psi}^{B} & = & [ \frac{1}{2}
({\hat p}^{AB^{\prime}} \ {\hat p}_{BB^{\prime}} - {\hat p}_{B}^{ \
B^{\prime }} \ {\hat p}^{A}_{ \ B^{\prime}} ) + \frac{1}{2} ({\hat p}%
^{AB^{\prime}} \ {\hat p}_{BB^{\prime}} + {\hat p}_{B}^{ \ B^{\prime}} \ {%
\hat p}^{A}_{ \ B^{\prime}} )] \ {\Psi}^{B}   \nonumber  \\
 & = & \frac{m^{2}}{2} \ {\Psi}^{A}.    \label{Dirac5}
\end{eqnarray}

\vskip10pt \noindent After some spinor manipulations, using the Fierz
identity in (\ref{Fierz}) and the definition in (\ref{mechpop}), we obtain

\begin{equation}
[{\hat p}_{KK^{\prime}} {\hat p}^{KK^{\prime}}
\delta^{B^{\prime}}_{A^{\prime }} + ie \phi_{A^{\prime}} ^{\ B^{\prime}}
]\Phi_{B^{\prime}} = m^{2} \Phi_{A^{\prime}},          \label{Dirac6}
\end{equation}

\begin{equation}
[{\hat p}^{KK^{\prime}} {\hat p}_{KK^{\prime}} \delta^{A}_{B} - ie {\bar\phi 
}^{A}_{\ B}] {\Psi}^{B} = m^{2} {\Psi}^{A},     \label{Dirac7}
\end{equation}

\vskip10pt \noindent where

\begin{equation}
\phi_{A^{\prime} B^{\prime}} :=\frac{1}{2} {\epsilon}^{AB} F_{AA^{\prime
}BB^{\prime}}=\phi_{B^{\prime}A^{\prime}} \ \ \text{and} \ \ {\bar\phi}_{AB}
:=\frac{1}{2} {\bar\epsilon}^{A^{\prime} B^{\prime}}
F_{AA^{\prime}BB^{\prime }}={\bar\phi}_{BA},    \label{phi}
\end{equation}

\vskip10pt \noindent define infinitesimal ${\overline{\mathrm{SL}(2,\C)}}$ and $\mathrm{SL}(2,\C)
$ transformations, i.e.\@ infinitesimal ``Lorentz'' transformations of the
two spinors at each space-time point. The electro-magnetic field $F$
is defined by the four-potential introduced in (\ref{A}) according to the
familiar rule

\begin{equation}
F_{AA^{\prime}BB^{\prime}} :=\partial_{AA^{\prime}} A _{BB^{\prime}}
-\partial_{BB^{\prime}} A_{AA^{\prime}}.   \label{F}
\end{equation}

\vskip10pt \noindent For the purpose of physical interpretation we express
the electro-magnetic Lorentz tensor field as

\begin{equation}
F_{AA^{\prime}BB^{\prime}}:= F_{ab}\Sigma^{ab}_{AA^{\prime}BB^{\prime}} ,  \label{Sigmaop}
\end{equation}

\vskip10pt \noindent where $(\Sigma^{ab})^{AA^{\prime}}_{\ \ BB^{\prime}}$
represents six infinitesimal generators of the Lorentz group acting on the
space of mixed spinor-tensors of rank two. However, this generators are
composed of two sets of six infinitesimal generators acting on the complex
vector space of simple spinors and their complex conjugates. This can be
seen very easily because in the usual spinor algebra manner we obtain the
decomposition

\begin{equation}
\Sigma^{ab}_{AA^{\prime}BB^{\prime}} := {\bar\Sigma}^{ab}_{ \ A^{\prime
}B^{\prime}} \epsilon_{AB} + \Sigma^{ab}_{\ AB} {\bar\epsilon}_{A^{\prime
}B^{\prime}},   \label{Sigmaopdef}
\end{equation}

\vskip10pt \noindent where the first set of six infinitesimal ${\overline
{\mathrm{SL}(2,\C)}}$ transformations (essentially the same as relativistic
spin-operator action on $\bar S$ (\ref{spinorspace})) and the second set of
six infinitesimal $\mathrm{SL}(2,\C)$ transformations (essentially the same
relativistic spin-operator action on $S$ in (\ref{spinorspace})) are given
by

\begin{equation}
\Sigma^{ab}_{\ AB}:= {\bar\epsilon}^{A^{\prime}B^{\prime}} \sigma
^{[a}_{AA^{\prime}} \sigma^{b]}_{BB^{\prime}}, \ \ \ {\bar\Sigma}^{ab}_{\
A^{\prime}B^{\prime}}:= {\epsilon}^{AB}
\sigma^{[a}_{AA^{\prime}}\sigma^{b]}_{BB^{\prime}},   \label{Diracsigma}
\end{equation}

\vskip10pt \noindent with $\sigma^{a}_{AA^{\prime}}$ defining the
isomorphism between Lorentz four-vectors and mixed spinor-tensors of second
rank. The two dimensional Dirac matrices $\sigma^{a}_{AA^{\prime}}$ are\
also called Infeld-Van der Waerden connecting quantities.
This has been explained at the end of section two in this review. We thus
have that

\begin{equation}
ie\phi_{A^{\prime}}^{\ B^{\prime} }:=\frac{ie}{2} F_{ab}{\bar\Sigma}^{ab \ {%
B^{\prime}}}_{\ A^{\prime}} \ \ \text{and} \ \ -ie{\bar\phi}^{A}_{\ B}:=%
\frac{-ie}{2} F_{ab}\Sigma^{ab A}_{\ \ \ \ B },   \label{phi1}
\end{equation}

\vskip10pt \noindent which implies that at each space-time point the
electro-magnetic field $F$ defines the six infinitesimal parameters of the
infinitesimal ${\overline{\mathrm{SL}(2,\C)}}$ and $\mathrm{SL}(2,\C)$ transformations acting on
each of the two spinors $\Phi$ and $\Psi$ separately.

\vskip10pt \noindent Quantum mechanical operators, i.e.\@ essentially
infinitesimal generators of relevant algebras, in the classical limit,
become functions on apropriate phase spaces. For example, consider the
Klein-Gordon operator

\begin{equation}
{\hat H}:={\hat p}^{a}{\hat p}_{a},  \label{KGmhat}
\end{equation}

\vskip10pt \noindent with ${\hat p}_{a}$ defined in (\ref{mechpop}). In the
classical limit, (\ref{KGmhat}) may be regarded as a Poincar{\'e} scalar
function on the cotangent bundle of the affine Minkowski space $T^{*}M$,

\begin{equation}
\mathcal{H}(P_{b},\ x^{a}):=({P}_{b}-eA_{b}(x^{a}))({P}^{b}-eA^{b}(x^{a})). \label{KGm}
\end{equation}

\vskip10pt \noindent The natural Poincar{\'e} invariant symplectic structure
on $T^{*}M$ is then defined by the only non-vanishing set of Poisson
brackets

\begin{equation}
\{P_{b}, \ x^{a} \}= \delta^{a}_{b},  \label{PX}
\end{equation}

\vskip10pt \noindent where $P_{b}$ and $x^{a}$ are the global Poincar{%
\'e} covariant/invariant canonical coordinates on $T^{*}M$. As is well-known,
\cite{ab1}, by taking the function in (\ref{KGm}) as a generator
of the canonical flow and projecting this flow onto the Minkowski base space
reproduces exactly the Lorentz force equation. The curves of this flow
projected onto the Minkowski base space are solutions of the Lorentz force
equation.

\vskip10pt \noindent Encouraged by this fact we were curious about what kind of
dynamical relativistic classical equations would appear from such a
dequantization of the Dirac equation minimally coupled with an external
electro-magnetic field. In other words, we wanted to see what kind of
modifications to the Lorentz force equation would be introduced by the classical\
limit of the electron-like spin (gyromagnetic ratio is automatically equal to 2\
for an electron in the Dirac equation). We have now, at our disposal,
the classical two-twistor phase space $T \Delta T$. With reference
to our discussion above, the classical limit of the six generators of the
infinitesimal ${\overline{\mathrm{SL}(2,\C)}}$ and $\mathrm{SL}(2,\C)$
transformations should have the intrinsic angular four-momentum function\
(\ref{Sigma}) on $T \Delta T$ as its classical limit. We make it to our\
assertion and define the flow generating function on the phase space\
$T\Delta T$ to be

\begin{equation}
H=H(Z,\ W, \ {\bar Z}, \ {\bar W})=(P_{i}-eA_{i})(P^{i}-eA^{i})+{\frac{1}{2}}%
e{\Sigma}_{kl}F^{kl},   \label{H}
\end{equation}

\vskip10pt \noindent which may be regarded as a classical limit of the
minimally coupled second order Dirac operator in (\ref{Dirac6}) and (\ref
{Dirac7}). If spin tensor $\Sigma$ vanishes, i.e. when the conformal scalars $%
\rho$ in (\ref{ZWscalar}) and $k$ in (\ref{k}) are equal to zero, the
function in (\ref{H}) coincides with (\ref{KGm}) and the decomposition of $%
T\Delta T$ is reduced to the direct sum $T^{*}M \oplus T^{*}S^{1}$. The flow
generated in this simplified two-twistor space reproduces again the Lorentz
force equation. However, there is a small, and maybe welcome, difference
here: the charge comes in as a constant of motion and not just as a number
put in by hand. This is a further justification of the definition in (\ref{H}%
).

\vskip10pt \noindent On the sixteen dimensional phase space $T\Delta T$ we
are only interested in the dynamics of the Poincar{\'e} invariant/covariant
functions $x, P, \Sigma$ and $e$. This makes fourteen variables altogether.
There are two additional angle variables, the argument of the Poincar{\'e}
scalar function $f$ in $(\ref{f})$ (the absolute value of $f$ is equal to
the Minkowski time-like length of the canonical four-momentum $P$ divided by 
$\sqrt{2} $ ) which is canonically conjugate to the charge $e$ and also the
phase (the flag) of the $\eta$ spinor canonically conjugate to the conformal
real valued scalar $k$ in ($\ref{k}$). These facts have been discoverd in 
\cite{zab,zab1}. The generating function in (\ref{H}) is such that the
dynamics of these two angles do not mix with the dynamics of the
``physical'' variables $x, P, \Sigma, e$ and therefore will not be displayed
in what follows. The canonical flow\footnote{%
it is canonical with respect to the symplectic structure defined by the
imaginary part of the pseudo-hermitian form preserved by the $SU(2,2)$
transformations.} on the phase space $T\Delta T$ generated by the Poincar{%
\'e} scalar function in (\ref{H}) implies the following equations for the
physical variables $x, P, \Sigma, e$:

\begin{equation}
\frac{d\ e}{d\lambda}:=\{H, e\}=0,  \label{edotlambda}
\end{equation}

\begin{equation}
\frac{dx^{j}}{d\lambda}:=\{H,\ x^{j}\}=2(P^{j}-eA^{j}), \label{xdotlambda}
\end{equation}

\begin{equation}
\frac{dP_{j}}{d\lambda}:= \{H,\ P_{j}\}=2e(P_{i}-eA_{i})\frac{\partial A^{i}%
}{\partial x^{j}} -{\frac{1}{2}}e{\Sigma}_{kl}\frac{\partial F^{kl}}{%
\partial x^{j}},    \label{Pdotlambda}
\end{equation}

\begin{equation*}
\frac{d{\Sigma}^{ij}}{d\lambda} := \{H,\, {\Sigma}^{ij}\}={\frac{1}{2}}%
eF^{}_{kl}\{{\Sigma}^{kl},\ {\Sigma}^{ij}\} = 
\end{equation*}

\begin{equation*}
={\frac{1}{2}}eF_{kl}(g^{ik}{\Sigma}^{lj}- g^{il}{\Sigma}^{kj}+g^{jl}{\Sigma 
}^{ki}-g^{jk}{\Sigma}^{li}) = 
\end{equation*}

\begin{equation}
={\frac{1}{2}}e(F^{i}\ _{l}{\Sigma}^{lj}- F_{k}\ ^{i}{\Sigma}^{kj}+F_{k}\
^{j}{\Sigma}^{ki}- F^{j}\ _{l}{\Sigma}^{li}),   \label{Sigmadotlambda}
\end{equation}

\vskip10pt \noindent where $\lambda$ labels points along the lines of the
flow. It would be more illuminating to have the proper time of the
two-twistor particle as an evolution parameter along the flow instead of $%
\lambda$. For notational reasons we first put

\begin{equation}
{\Sigma}F:={\frac{1}{2}}\Sigma^{kl}F^{}_{kl},  \label{SF}
\end{equation}

\vskip10pt \noindent and note that the relation between the proper time $\tau$
and the parameter $\lambda$ reads

\begin{equation}
(\frac{d\tau}{d\lambda})^{2}:= \frac{dx^{j}}{d\lambda} \frac{dx_{j}}{d\lambda%
} = 4(P^{j}-eA^{j})(P_{j}-eA_{j})=4(H-e\Sigma F)^{2},  \label{propertime}
\end{equation}

\vskip10pt \noindent where $H$ denotes the constant value of the generating
function (trivially, it is a constant of (the flow) motion) that can be
identified with the dynamical mass squared of the system. The equation in (%
\ref{propertime}) was obtained from (\ref{xdotlambda}) with additional use
of (\ref{H}) where the mechanical four-momentum squared is the difference
between the value of $H$ (the dynamical mass squared of the system) and the
contribution to this value coming from the interaction of the spinning
charge with the external electro-magnetic field. If the dot denotes
differentiation with respect to the proper time

\begin{equation}
{\dot x}^{j} := \frac{dx^{j}}{d\tau} =\frac{P^{j}-eA^{j}}{\sqrt{H-e{\Sigma }%
F^{}}},   \label{xdottau}
\end{equation}

\vskip10pt \noindent then we obtain

\begin{equation}
{\dot x}^{j} {\dot x}_{j}=1.  \label{4velocity}
\end{equation}

\vskip10pt \noindent as it should be.

\vskip10pt \noindent Eliminating the equation of motion for the canonical
linear four-momentum, we get finally

\begin{equation}
\frac{d e}{d\tau}=0,   \label{edottau}
\end{equation}

\begin{equation}
\frac{d^{2} x^{j}}{d{\tau}^{2}} = \frac{e}{\sqrt{H-e{\Sigma}F^{}}} \ F_{i}\
^{j}\ \frac{dx^{i}}{d{\tau}}+ \\
\frac{e}{4 (H-e{\Sigma}F^{})}[\frac{\partial{F^{}_{kl}}}{\partial{x^{m}}} {%
\Sigma}^{kl} (\frac{dx^{m}}{d{\tau}} \frac{dx^{j}}{d{\tau}} - g^{mj})],  \label{Bette}
\end{equation}

\begin{equation}
\frac{d{\Sigma}^{ij}}{d{\tau}} =\frac{e}{4 \sqrt{H-e{\Sigma}F^{}}} (F^{i}\
_{l}{\Sigma}^{lj}- F_{k}\ ^{i}{\Sigma}^{kj}+F_{k}\ ^{j}{\Sigma}^{ki}-F^{j}\
_{l}{\Sigma}^{li}).   \label{Sigmadottau}
\end{equation}

\vskip10pt \noindent The equation in (\ref{Bette}) gives a
generalisation of the Lorentz force equation following from the two-twistor
dequantization of the minimally coupled Dirac equation. The equation of
motion for the (five\footnote{%
one of the Lorentz invariant scalars formed from $\Sigma$ equals zero, as
mentioned before.}) spin variables in (\ref{Sigmadottau}), automatically
implied by the formalism, differs significantly from the so called BMT
(Bargmann, Michel, Telegdi) equation. In the latter, the number of spin
variables equals three. Besides, one also requires the norm of the spin
four-vector to be a constant of motion. Consequently there are, in
total, only two independent variables. Such spin variables have the origin in an, a
priori, defined Pauli-Luba{\'n}ski spin four-vector $S^{a}$ which is
constructed as a relativistic generalisation of the non-relativistic spin
vector \cite{Jackson,Rohrlich}. The so defined $S^{a}$ has also to fulfill a
constraint\footnote{%
very hard if not, in general, impossible to fulfil consistenly with the
condition that the particle is point-like and described by a Hamiltonian
principle.}, $S_{a}{\dot x^{a}}=0$, and the requirement that $-S_{a}{S^{a}}$
is a constant of motion, as was mentioned above.

\vskip10pt \noindent To our knowledge, the so defined classical BMT equation
together with (a number of different suggested versions trying to
generalise) the Lorentz force equation so that it also includes spin
variables, have never been given a proper relativistic hamiltonian
description. The starting point has always been non-relativistic classical
mechanics. The relevant discussions concerning these matters may be found in
Jackson's, Corben's and Rohrlich's books \cite{Jackson,Corben,Rohrlich}.
There exists also a number of Lagrangian formulations. Some of them make use
of the so called anticommuting fermionic numbers and Grassman variables. A
relatively recent resume of Lagrangian formulations may be found in \cite
{Frydyszak}.

\vskip10pt \noindent To get our relativistic Hamiltonian formulation, we
start with the relativistic Dirac quantum mechanical equation (at the first
quantisation level), ``classicalise'' it on the $T \Delta T$ and thereby
obtain the equations in (\ref{edottau}), (\ref{Bette}), (\ref{Sigmadottau}).
The relativistic Hamiltonian formulation is already there.

\vskip10pt \noindent Note that from (\ref{Sigmadottau}) it also
automatically follows that the conformal scalar defined in (\ref{absvalue})
is a constant of motion.

\vskip10pt \noindent We do not insist on the fact that the particle should be
point-like. On the contrary, for a free two-twistor object two
four-positions (one commuting and one non-commuting) were distinguished. The
commuting $x$ in (\ref{4pos}) was identified with the centre of charge while
the non-commuting one $q$ in (\ref{4poswithspin}) was identified with the
centre of mass. With the external electro-magnetic field switched on the
four position $x$ plays the role of the dynamical variable. The
canonical linear momentum four-vector $P$ and the ``canonical''
four-position of the center of mass $q$ in (\ref{4poswithspin}) are totally
eliminated from the equations of motion. We suggest that the Lorentz
space-like four-vector pointing from the centre of charge to a new dynamical
centre of mass of the system (now interacting with an external
electro-magnetic field) should be defined by

\begin{equation}
\Delta q^{i} :=\frac{1}{\sqrt H} {\Sigma}^{ik}{\dot x}_{k},  \label{Deltax}
\end{equation}

\vskip10pt \noindent This coincides with $\Delta x^{a}$ in (\ref{Deltapos})
in the case when the external electro-magnetic field is zero, i. e. $F_{ab}=0$.\
The new dynamical intrinsic angular four-momentum with respect to the centre of
charge then reads

\begin{equation}
S^{ij}:= {\Sigma}^{ij} - \sqrt{H}({\dot x}^{i} \Delta q^{j} - {\dot x}%
^{j}\Delta q^{i} )={\Sigma}^{ij}- {\dot x}^{i} {\dot x}_{k}{\Sigma}^{jk} + {%
\dot x}^{j} {\dot x}_{k}{\Sigma}^{ik},   \label{Sij}
\end{equation}

\vskip10pt \noindent and coincides with $S^{ij}$ in (\ref{orbital+Lubanski})
when $F_{ab}=0$. The new dynamical Pauli-Luba{\'n}ski spin four-vector $S^{i}$\
reads now as

\begin{equation}
{S^{i}}:= \frac{\sqrt{H}}{2} \epsilon^{ijkl}{\Sigma}_{jk}{\dot x}_{l},  \label{Si}
\end{equation}

\vskip10pt \noindent and coincides with $S^{ij}$ in (\ref{Pauli-Lubanski}) when
$F_{ab}=0$.

\vskip10pt \noindent The value of the square of the Lorentz norm of this
newly introduced dynamical spin four-vector ($S$ and $\Delta q$ are
space-like Lorentz four-vectors hence the negative signs) is then given by

\begin{equation}
S^{2}:=\frac{- S_{i}S^{i}}{H} = k^{2} - {\Delta q^{i}\Delta q_{i}}{H}, 
                                               \label{Paulilubanskispin}
\end{equation}

\vskip10pt \noindent and, unlike the value of the function $%
\Sigma^{ab}\Sigma_{ab}=4k^{2}$, is not a constant of motion (except perhaps
for some special choices of the external electro-magnetic field); instead we
have

\begin{equation}
\frac{dS^{2}}{d\tau}=-H\frac{d{(\Delta q^{i}\Delta q_{i})}}{d\tau} \label{Spinderivative}
\end{equation}

\vskip10pt \noindent showing that the value of the square of the norm of the
intrinsic (intrinsic with respect to the centre of charge) spin four-vector
varies as the square of the distance between the centre of mass and the
centre of charge multiplied by the square of the mass of the system. >From (%
\ref{Paulilubanskispin}) and (\ref{Spinderivative}) it now follows that if
the Pauli-Luba{\'n}ski spin decreases, then the distance between the centre of
charge and centre of mass increases and vice versa. Is this a new classical
interpretation of the famous ``Zitterbewegung''?

\vskip10pt \noindent Let us now take a look on how the dynamics described above
looks like when it is formulated as a minimal action principle on $T \Delta T$.
This will take us closer to the, as yet unknown, special relativistic
twistor quantum dynamics.

\section{AN ACTION PRINCIPLE ON THE TWO-TWISTOR SPACE.}

We start with the derivation of the Lorentz force equation from a Poincar{\'e%
} invariant action principle not on the two-twistor space but on $T^{*}M$.

\vskip10pt \noindent The Poincar{\'e} invariant symplectic potential
defining Poisson bracket structure (which in terms of global canonical
Poincar{\'e} covariant variables was defined by the canonical Poisson
bracket relations in (\ref{PX})) on $T^{*}M$, the (8D) cotangent bundle of
the Minkowski space time, is given by\footnote{$T^{*}M$ equipped with the
symplectic structure $\Omega_{0}$ defined by $\gamma_{0}$ ($%
\Omega_{0}=d\gamma_{0}$) is called the extended phase space of a spinless
particle; extended because the Poincar{\'e} group acts \emph{non}-transitively\
on $T^{*}M$, transitivity being the classical analog of irreducibility.}

\begin{equation}
\gamma_{0}:=P_{i}dx^{i},   \label{gamma0}
\end{equation}

\vskip10pt \noindent where $P_{i}$ denotes the coordinates of a Lorentz four
(co-) vector (i.e.\@ a vector in $\R^{4}$ regarded as a covariant tensor of
rank one with respect to the Lorentz group), while $x^{i}$ denotes the\
coordinates of a Minkowski position four vector (i.e.\@ an affine vector in $%
\R^{4}$ regarded as a contravariant Lorentz four-vector of rank one).

\vskip10pt \noindent Consider extremal curves (with fixed endpoints) in $%
T^{*}M$ of the functional

\begin{equation}
{\tilde S}({\tilde C}):=\int_{{\tilde C}}{\gamma_{0}},  \label{Stilde}
\end{equation}

\vskip10pt \noindent where all the curves $\{ \tilde C \}$ are constrained
to lie on the (7D) hypersurface in $T^{*}M$ defined by a Poincar{\'e}
invariant condition

\begin{equation}
{\tilde C} \subset\{(x, \ P) \ \epsilon\ T^{*}M; \
[P_{i}-eA_{i}(x^{k})][P^{i}-eA^{i}(x^{k})]-\mathcal{H} =0 \},  \label{Lorentz}
\end{equation}

\vskip10pt \noindent with $m^{2}:=\mathcal{H}>0$ and $e$ being non-zero
constants and with $A_{i}(x^{k})$ being a real Lorentz four (co-) vector
valued function on $M$.

\vskip10pt \noindent As is well-known, projections of these extremal curves
onto the Minkowski space $M$ give space-time trajectories of a charged
massive (non-spinning) particle, with fixed charge $e$ and fixed mass $m$,
moving under the action of an external electro-magnetic field defined by the
four-potential $A_{i}(x^{k})$. These trajectories are solutions of the so
called "Lorentz force equation". The Lorentz force equation itself may also
be derived from this principle.

\vskip10pt \noindent To derive it explicitly and thereby to prove our
assertion, one replaces the action in (\ref{Stilde}), subject to the
condition in (\ref{Lorentz}), by a new action (Lagrange's multiplier method)

\begin{equation}
S(C):=\int_{C}\{{\gamma_{0}} - \frac{1}{2} l(\lambda)(
[P_{i}-eA_{i}(x^{k})][P^{i}-eA^{i}(x^{k})]-m^{2})d\lambda\} ,  \label{S}
\end{equation}

\vskip10pt \noindent where one lets $l(\lambda)$ to be an arbitrary real
valued function (Lagrange multiplier to be varied over) of an arbitrary real
valued parameter $\lambda$ labeling points along $C$, which is now allowed
to lie anywhere in $T^{*}M$ (but with endpoints still fixed at the same
positions on the surface defined in (\ref{Lorentz})). The extremal curves of
this new action coincide with the extremal curves of the action in (\ref
{Stilde}) subject to the condition given by (\ref{Lorentz}). Thus, we look
after the extremal curves of the functional

\begin{equation}
S(C(\lambda) ):=\int_{C} L (C(\lambda)) d{\lambda} ,  \label{S1}
\end{equation}

\vskip10pt \noindent where (the Lagrangian) $L(C(\lambda))$ on $T^{*}M$ is
given by

\begin{equation}
L(C(\lambda))=P_{i}(\lambda) \frac{dx^{i}}{d\lambda} - \frac{1}{2}
l(\lambda)(
[P_{i}(\lambda)-eA_{i}(x^{k}(\lambda))][P^{i}(\lambda)-eA^{i}(x^{k}(%
\lambda))]-m^{2}).        \label{Lagrange}
\end{equation}

\vskip10pt \noindent Here $P_{i}(\lambda)$, $x^{i}(\lambda)$ and $l(\lambda)$ in (%
\ref{Lagrange}) are regarded as independent (function-) variables of the
functional $S$. Our point is to look for the extremum of $S$. First, we note
that in (\ref{Lagrange}) there are no derivatives (with respect to $\lambda$%
) of the functions $P_{i}(\lambda)$ and of the function $l(\lambda)$. Therefore,
the Euler-Lagrange equations with respect to $P_{i}(\lambda)$ give

\begin{equation}
\frac{dx_{i}}{d\lambda} - l(\lambda)
[P_{i}(\lambda)-eA_{i}(x^{k}(\lambda))]=0,  \label{P}
\end{equation}

\vskip10pt \noindent or, equivalently,

\begin{equation}
P_{i}(\lambda)=\frac{1}{l(\lambda)} \frac{dx_{i}}{d\lambda} + e
A_{i}(x^{k}(\lambda)).  \label{P1}
\end{equation}

\vskip10pt \noindent When we plug back this equatoin into (\ref{Lagrange}) (in this way
eliminating the variable $P_{i}(\lambda)$ from (\ref{Lagrange})) yields

\begin{equation}
L(C(\lambda))=(\frac{1}{l(\lambda) } \frac{dx_{i}}{d\lambda}+ e
A_{i}(x^{k}(\lambda))) \frac{dx^{i}}{d\lambda} - \frac{1}{ 2} l(\lambda)(
\frac {1}{(l(\lambda))^{2}} \frac{dx^{i}}{d\lambda} \frac{dx_{i}}{d\lambda}
-m^{2}),           \label{L3}
\end{equation}

\vskip10pt \noindent or equivalently,

\begin{equation}
L(C(\lambda))={\frac{1}{2}}( \frac{1}{l(\lambda) } \frac{dx_{i}}{d\lambda} 
\frac{dx^{i}}{d\lambda} +l(\lambda)m^{2}) + e A_{i}(x^{k}(\lambda)) \frac{%
dx^{i}}{d\lambda}.  \label{L4}
\end{equation}

\vskip10pt \noindent The Euler-Lagrange equations with respect to $l(\lambda)$
applied to (\ref{L4}) give

\begin{equation}
{\frac{1}{2}}(- \frac{1}{[l(\lambda)]^{2}} \frac{dx_{i}}{d\lambda} \frac{%
dx^{i}}{d\lambda} + m^{2})=0 ,  \label{L5}
\end{equation}

\vskip10pt \noindent which yields

\begin{equation}
l(\lambda)= \pm\frac{\sqrt{\frac{dx_{i}}{d\lambda} \frac{dx^{i}}{d\lambda}}}{
m}.   \label{L6}
\end{equation}

\vskip10pt \noindent By plugging back (\ref{L6}) into (\ref{L4}) (and
thereby eliminating the variable $l(\lambda)$ from (\ref{L4})), we obtain a
Lagrangian defined entirely on the (base=configuration) Minkowski space:

\begin{equation}
L(C(\lambda)) =\pm m \sqrt{\frac{dx^{i}}{d\lambda} \frac{dx_{i}}{d\lambda}}%
+e A_{i}(x^{m}(\lambda)) \frac{dx^{i}}{d\lambda}.   \label{L1}
\end{equation}

\vskip10pt \noindent The Lagrangian $L(C(\lambda))$ in (\ref{L1}) is, of course (up to the
sign ambiguity), the familiar Lagrangian one usually starts with. Its
Euler-Lagrange equations, extremalizing the projection (onto the Minkowski
space) of the original action given in (\ref{Stilde}) subject to the
condition in (\ref{Lorentz}) or equivalently, as given in (\ref{S}),
constitute the well-known Lorentz force equation if (in the so arising
equations) one chooses the parameter $\lambda$ to be the proper time of the
particle, i.e.\@ if, in the final equation, one requires that $\frac{dx^{i}}{%
d\lambda} \frac{dx_{i}}{d\lambda}=1$.

\vskip10pt \noindent Concluding we note that the action principle defined
above in (\ref{Stilde}) and (\ref{Lorentz}) is equivalent to the method
described in the previous section. Therefore we now turn our attention back
to the two-twistor phase space $T \Delta T$.

\vskip10pt \noindent On $T \Delta T$, the (16D) two-twistor space, consider
a global symplectic potential (one-form), which is invariant with respect to 
$SU(2, 2)$ (therefore also essentially\footnote{%
this word indicates that only identity component of the universal
covering group of the Poincar{\'e} group is considered.} with respect to the
Poincar{\'e} group), given by $\gamma:=\gamma_{1}+\gamma_{2} $, where
\begin{equation}
\gamma_{1}:= \frac{1}{2}(iZ^{\alpha}d{\bar Z}_{\alpha}-i{\bar Z}_{\alpha
}dZ^{\alpha}) \ \text{and} \ \gamma_{2}:=\frac{1}{2}(iW^{\alpha}d{\bar W}%
_{\alpha}-i{\bar W}_{\alpha}dW^{\alpha}).   \label{gamma}
\end{equation}

\vskip10pt \noindent We remind that the symplectic
structure $\Omega_{i} := d{\gamma}_{i}$ defined by $\gamma_{1}$ or $\gamma
_{2}$ in (\ref{gamma}) coincides with the imaginary part of the hermitian
form (scalar "product") which is preserved by the $SU(2,2)$ (and therefore
(essentially) by the Poincar{\'e}) group acting on the twistor space $T$. In
effect the Poisson bracket structure defined by $\Omega:= d{\gamma}$ is
conformally (and therefore essentially also Poincar{\'e}) invariant. In
other words, the conformal transformations of $T \Delta T$ are canonical with
respect to the symplectic structure $\Omega$. In the previous section we have
already used this symplectic structure at the level of the corresponding
conformally invariant Poisson bracket algebra.

\vskip10pt \noindent In the previous section we have proved
that the symplectic structure on $T \Delta T$ is decomposed into three Poincar{\'e}
invariant parts. The Poincar{\'e} invariant form of
the symplectic potential in (\ref{gamma}) is then written as follows

\begin{equation}
\gamma= P_{i}dx^{i} + ([{\sigma}^{A^{\prime}}] d[\eta_{A^{\prime}}] +[{%
\bar\sigma}^{A}]d{[\bar\eta}_{A}]) + e d{\phi} .   \label{+}
\end{equation}

\vskip10pt \noindent The pair of square bracketed spinor cooordinates $%
([\eta_{A^{\prime}}], \ [{\sigma}^{A^{\prime}}])$ on the real projective
spinor space $T^{*} \R P({S})$ is intended to recall us that these are just
equivalence classes with respect to the multiplication (division) by a
non zero real number $r$ according to the rule \cite{zab,zab1}:

\begin{equation*}
(\eta_{A^{\prime}}, \ {\sigma}^{A^{\prime}}) \equiv(r \ \eta_{A^{\prime}}, \ 
\frac{1}{r}\ {\sigma}^{A^{\prime}}).
\end{equation*}

\vskip10pt \noindent Recall that $P_{a}$, $x^{a}$, ${\sigma}^{A^{\prime}}$, $\eta
_{A^{\prime}}$, $e$ and $\phi=\arg{f}$ in ($\ref{+}$) were defined in (\ref
{Poincare1+2P}), (\ref{4pos}), (\ref{sigm}), (\ref{ZWomegapilambdaeta}), (%
\ref{e}) and (\ref{f}). Inserting carefully this chain of definitions into (%
\ref{+}) and finally using the spinor representation of the two-twistors as
in (\ref{ZWomegapilambdaeta}) reproduce the symplectic potential in (\ref
{gamma}). The reader should perhaps try to perform this spinor algebra
manipulations just to convince himself or perhaps find some sign
inconsitencies that must be corrected.

\vskip10pt \noindent Consider now extremal curves $\{\tilde C\}$ (with fixed
endpoints) in $T\Delta T$ of the functional

\begin{equation}
{\tilde S}_{spin}({\tilde C}):=\int_{{\tilde C}}{\gamma},   \label{Stildetwist}
\end{equation}

\vskip10pt \noindent where all these curves $\{\tilde C\}$ are required to
lie on a (15D) hypersurface in $T\Delta T$ defined by the condition

\begin{equation}
[P_{i}-eA_{i}(x^{k})][P^{i}-eA^{i}(x^{k})]+\frac{1}{2} e
\Sigma^{ij}F_{ij}(x^{k})=H.    \label{Aspin}
\end{equation}

\vskip10pt \noindent Here $H=\text{const}$, $A_{i}(x^{a})$, as usual,
represents the (real) four-potential of an external electro-magnetic field $%
F_{ij}(x^{k})=-F_{ji}(x^{k})$ acting on the charged massive and spinning
particle with its charge equal to the value of function $e$ defined in (\ref
{e}) which is a constant of motion.

\vskip10pt \noindent Without any proof we can now claim that the equations
of motion implied by this action principle are exactly the ones described in
the previous section in (\ref{edottau}), (\ref{Bette}), (\ref{Sigmadottau}).
However, this is much harder to prove directly using the action principle in (%
\ref{Stildetwist}), (\ref{Aspin}) because the symplectic potential contains
the $\sigma$ and $\eta$ variables so that they will appear explicitly in the
arising equations of motion while we really wish that only $\Sigma$
variables should appear explicitly (and $P$, $x$ and $e$) in the arising
equations of motion.

\vskip10pt \noindent The equations of motion obtained in the previous
section must therefore be constructed a posteriori. By the equivalence of
the two aproaches we know without any calculations that the arising
equations for the variables $P$, $x$, $e$ and $\Sigma$ are the same as those
in (\ref{edotlambda}), (\ref{xdotlambda}), (\ref{Pdotlambda}) and (\ref
{Sigmadotlambda}).

\vskip10pt \noindent At the quantum level one is not interested in the
equations of motion. The entire action integral is used in quantum
mechanics. All paths between two fixed points on the fifteen dimensional
surface are of importance, not just the extremal curves. Therefore the above
action integral formulation could serve as a starting point to a fresh
quantization of the dynamics of a massive, charged and spinning particle
thereby generalizing the minimally coupled Dirac equation. The arising
equations would then be valid for any quantised value of the spin. Stopping
here, we hope that we will be able to come back to this issue in the nearest
future.

\section{CONCLUSIONS AND REMARKS.}

The results presented in this review are not in the main stream of the
research within the Twistor Theory. The very ambitious goals of such a
research in this main stream led by Roger Penrose, concern the translation
of the general relativity theory into the language of holomorphic functions
of (many?) twistors. Penrose hopes that this approach will lead to a
unification of quantum field theory with the general theory of relativity
(of course, describing gravitation in a curved space-time, no longer in the
affine Minkowski space) in a way very different from that in the research
pursued by the people working with the superstrings, M-theory, etc.\@.
Penrose thinks that both quantum principles and GR must be modified in some
way. This way, as Penrose put it, will be pointed out by the many
dimensional singularity sets of functions of many twistors and also by means
of sofisticated analysis and geometry in multidimensional Riemann-like
surfaces defined by these many complex variables holomorphic functions. The
singularity surfaces will, according to Penrose, replace the notion of the
quantum mechanical Hilbert space of states, simultaneously turning the wave
function to an (conformally invariant?)\@ objective entity. This could solve
problems with the understanding of the quantum mechanical process of
measurement. The gravity effect would then be responsible for the so called
collapse of the wave function. The quantum mechanical superposition
principle must then be abandoned (being only the first approximation of the
new formulation) for some new and not yet known, but more accurate, principles
should replace it. Up to now this goal has been achieved only partially 
\cite{QG,MathSoc,QG1}.

\vskip10pt \noindent In this review our goal was very much modest.
At the level of special relativistic classical particle dynamics we wanted
to see what new elements are brought in if the twistor description was
imposed. I think we succeeded quite well. The idea that everything is made
of non-local masslessness, inherent in the twistor formalism from the very
start by the built-in conformal symmetry, seems to be very appealing. The
holomorphic aspects are not so important at the level of classical
relativistic twistor dynamics. Instead, the symplectic structure (the imaginary
part of the pseudo-hermitian form preserved by $SU(2,2)$) is emerging as the
most important element of such a twistor approach. This is not so surprising
because a conformally invariant first quantization leads automatically to
the aspects of holomorphy \cite{pr3,prmc}.

\vskip10pt \noindent We think that the classical equations of motion of the
type described in this review are worthwhile to investigate. It would be
interesting to explore some possible consequences of the obtained equations
of motion above. For that reason we should reintroduce the Planck constant $%
\hbar$ and the velocity of light constant $c$ into the obtained equations
and investigate the non-relativistic limit to some order of $(\frac{\sqrt{{%
\dot x^{a}} {\dot x_{a}}}}{c})^{n} $ and some order of $(\frac{\hbar}{k})^{m}
$. Once this is done, it would be of outmost importance to recast the so
obtained approximations into a new non-relativistic hamiltonian formulation
containing relativistic and spin induced correction terms coming from our
approach. Such approximative models should have predictive power and perhaps
some experimental proposals could emerge to test them. These concrete
calculations are left for the future while we hope to be able to perform
them in cooperation with the members of the research group of Professor
Iouri Mikhailovich Vorobiev who are experts in this domain.

\section{ACKNOWLEDGMENTS.}

We wish to thank Professor Iouri Mikhailovich Vorobiev, Professor Rub\'en
Flores Espinoza and Guillermo D\'avila Rasc\'on, of Departmento de Matem\'aticas\
de la Universidad de Sonora, for giving me the opportunity, within the frame\
of the project ``ACOPLAMIENTO MINIMO EN LA GEOMETRIA DE POISSON Y LA TEORIA\
DE SISTEMAS HAMILTONIANOS'' sponsored by CONACYT with project number\
35212-E, to perform the research presented in this paper. I also wish to\
thank KTH-Syd (Royal Institute of Technology) for providing me with so the\
called ``fakir'' funds and thereby giving me time for this research by\
arranging, at the right moments, a leave of absence from my teaching duties.

\end{document}